\documentclass[review,12pt]{elsarticle} 
\usepackage[T1]{fontenc}              
\usepackage[utf8]{inputenc}           
\usepackage[english]{babel}           
\usepackage{lmodern}                  
\usepackage{microtype}                

\usepackage{amsmath,amssymb,amsfonts} 

\usepackage{graphicx} 
\usepackage{subfig} 
\usepackage{floatrow} 
\usepackage{booktabs} 
\usepackage{rotating} 
\usepackage{dblfloatfix}

\usepackage{hyperref} 
\usepackage{color} 

\usepackage[group-minimum-digits = 5, group-separator={,}, per-mode=symbol]{siunitx}

\usepackage{textcomp} 

\journal{Journal of non-Newtonian Fluid Mechanics}

\begin{document}

\begin{frontmatter}

\title{Sedimentation and migration dynamics of a spherical particle in an elastoviscoplastic fluid near a wall}

\author{Alie Abbasi Yazdi}
\ead{aliyeh.abbasiyadzi@unina.it}

\author{Gaetano D'Avino\corref{cor1}}
\ead{gaetano.davino@unina.it}

\address{Dipartimento di Ingegneria Chimica, dei Materiali e della Produzione Industriale, Universit$\grave{a}$ di Napoli Federico II, P.le~Tecchio~80, 80125 Napoli, Italy}

\cortext[cor1]{Corresponding author. Tel.: +39 0817682241}

\begin{abstract}

The sedimentation of a spherical particle in an elastoviscoplastic fluid in proximity of a flat wall is investigated by direct numerical simulations. The governing equations under inertialess conditions are solved by the finite element method with an Arbitrary Lagrangian-Eulerian formulation to manage the particle motion. The fluid is modeled with the Giesekus constitutive equation modified as proposed by Saramito (2007).

The sedimentation, migration, and angular velocities are computed as a function of the particle-wall distance for various Weissenberg and Bingham numbers. The presence of a yield stress reduces the settling velocity and reverses the migration direction as compared to the purely viscoelastic case. The effect of the confining wall on the yielded and unyielded regions around the particle is investigated. The reversed particle migration phenomenon observed in the elastoviscoplastic fluid is attributed to the different shear rate distribution around the particle due to the presence of the yielded region. The negative wake behind the particle is discussed and related to the axial component of the viscoelastic stress.

\end{abstract}

\begin{keyword}
Elastoviscoplastic fluids \sep Sedimentation \sep Particle migration \sep Yield stress \sep Confinement  \sep Numerical simulations
\end{keyword}

\end{frontmatter}

\section{Introduction}
\label{sec:Introduction}

The transport of solid particles in non-Newtonian fluids is fundamental in a huge number of industrial and biological applications. The peculiar rheological properties of the suspending liquid are responsible for drastic variations of the particle dynamics as compared to Newtonian suspensions. A significant number of studies have focused on the effect of viscoelasticity on the motion of the suspended particles in a variety of flow fields \cite{DAvino2015particle}. A sphere sedimenting in a quiescent unconfined viscoelastic fluid, for instance, exhibits a variation of the steady-state drag correction factor (defined as the ratio between the drag coefficient of the particle in the viscoelastic fluid and in the Newtonian one) as the degree of elasticity is varied \cite{Mckinley2002transport}, and the appearance of the ‘negative wake’ behind the particle, i.e., a region downstream the sphere where the fluid velocity has direction opposite to the gravity \cite{Hassager1979negative,Harlen2002negative}. The presence of confining walls strongly affects both the drag coefficient and the negative wake phenomenon \cite{Mckinley2002transport}. Furthermore, a particle sufficiently close to a wall may also experience a motion orthogonal to the force direction referred as ‘lateral migration’ \cite{Singh2000sedimentation,Spanjaards2019numerical}. The migration phenomenon has been studied in other flow fields such as shear, Couette, channel flows, with the latter case of special interest for microfluidic applications \cite{DAvino2017particle}.

Another relevant property of non-Newtonian fluids is the existence of a ‘yield stress’, i.e., a threshold for the level of applied stress below or above which the material behaves like a solid or a liquid \cite{Bonn2009yield,Coussot2014yield}. Several works have investigated the settling dynamics of spherical particles in yield stress materials, also termed viscoplastic, both experimentally \cite{Atapattu1995creeping,Merkak2006spheres,Gueslin2006flow,Tabuteau2007drag,Putz2008settling,Mirzaagha2017rising,Sgreva2020interaction} and numerically \cite{Beris1985creeping,Blackery1997creeping,Beaulne1997creeping,Liu2002convergence,Wachs2016particle,Iglesias2020computing,Ferrari2021fully}. The results show that the drag correction factor increases as the yield stress increases, i.e., a sphere settles slower in a viscoplastic fluid rather than in a Newtonian fluid. Criteria for the stopping of the particle and the shape of the yielded/unyielded region as the relevant parameters are varied are also investigated. 

In a variety of real fluids, both elasticity and yield stress are present. These materials are referred as elastoviscoplastic (EVP) fluids. Elasticity must be considered in order to justify some phenomena observed in yield stress materials that are typical of viscoelastic fluids. This is the case of the loss of the fore–aft symmetry of the flow field around a settling particle under creeping conditions and the formation of the negative wake \cite{Gueslin2006flow,Putz2008settling}, both phenomena that are not predicted in viscoplastic fluids with negligible elasticity.

Despite their relevance, studies on the dynamics of particles in elastoviscoplastic fluids are relatively few. Fraggedakis et al. \cite{Fraggedakis2016yielding} incorporated elastic effects in a pure viscoplastic fluid using the constitutive equation proposed by Saramito \cite{Saramito2007new,Saramito2009new}. They solved the problem of a sphere sedimenting in a large container by a 2D axisymmetric finite element method. They found that typical phenomena such the negative wake and the loss of the fore-aft symmetry under creeping flow conditions are due to elasticity and not to thixotropy. The simulation results agree very well with the available experimental measurements \cite{Holenberg2012particle}. They also propose a relation for the drag correction coefficient as a function of the material plasticity for different levels of elasticity and the criteria for particle entrapment. The settling dynamics of a spherical particle in an uncofined EVP fluid has been later investigated by fully-resolved 3D numerical simulations in the absence and presence of a cross-shear flow orthogonal to the settling direction under weak inertia \cite{Sarabian2020numerical}. In the presence of a cross-shear flow, the drag on the sphere is reduced and the negative wake behind the particle disappears.

As previously mentioned, the presence of a confining wall strongly affects the particle dynamics giving rise to new phenomena such as the cross-stream lateral migration. It is interesting, then, to investigate the interplay between wall effect, fluid viscoelasticity and plasticity on the dynamics of particles. In this regards, Chaparian et al. \cite{Chaparian2020particle} numerically studied the dynamics of a neutrally buoyant rigid spherical particle in an elastoviscoplastic fluid in a pressure-driven flow between two infinite flat walls. The Saramito constitutive equation \cite{Saramito2007new} is employed. The Reynolds number is fixed to 20, the Weissenberg number is varied between 0.1 and 1, and the Bingham number between 0 and 0.4. They observed two sheared yielded regions near the walls and a core unyielded region. Particles initially released in the core region translate at the same velocity as the plug without migrating and rotating. On the other hand, the dynamics of the particles in the yielded regions depend on the relative weigh of elasticity and plasticity. In the viscoelastic limit, inertia is dominant and the particle migrates towards a position in between the centerline and the wall. In the viscoplastic limit, a similar behaviour is observed, with the yield surface taking the role of the centerline, so the migration occurs in the sheared regions only. Finally, when both elasticity and plasticity are relevant, for a certain range of Weissenberg numbers, the particle enters the plug region and moves towards the centreline.

More recently, Sarabian et al. \cite{Sarabian2022interface} investigated through numerical simulations the confinement effect on the sedimentation of a single spherical particle in elastoviscoplastic fluids between two parallel walls under weak inertial conditions. The fluid is modelled through the Saramito constitutive equation and the immersed boundary method is used to handle the rigid particle motion. The results show that, by increasing the confinement, the fluid velocity relaxes faster and the location of the negative wake moves towards the sphere, resulting in a lower sedimentation rate. The minimum value of the confinement ratio such that the wall effect becomes negligible and the maximum variation of the drag correction coefficient with confinement ratio are weakly dependent on the fluid elasticity and plasticity for a certain range of parameters. Also, they propose an expression of the drag coefficient as a function of the Bingham number and confinement ratio. It is important to point out that the particle is always initially located with center on the centerplane between the two walls, thus for symmetry does not migrate.

In this work, we study the dynamics of a spherical particle settling in an elastoviscoplastic fluid in proximity of a flat wall through direct numerical simulations under inertialess conditions. The sedimentation and migration phenomena are investigated by varying the particle-wall distance and the relevant dimensionless numbers. The yielded/unyielded fluid regions, the shear rate distribution around the particle, and the negative wake phenomenon are investigated and related to the observed dynamics. A comparison with the purely viscoelastic case is also discussed.

\section{Mathematical model}
\label{sec:Mathematical model}

\subsection{Governing equations}

\begin{figure}[t!]
\centering
\includegraphics[width=1.0\textwidth]{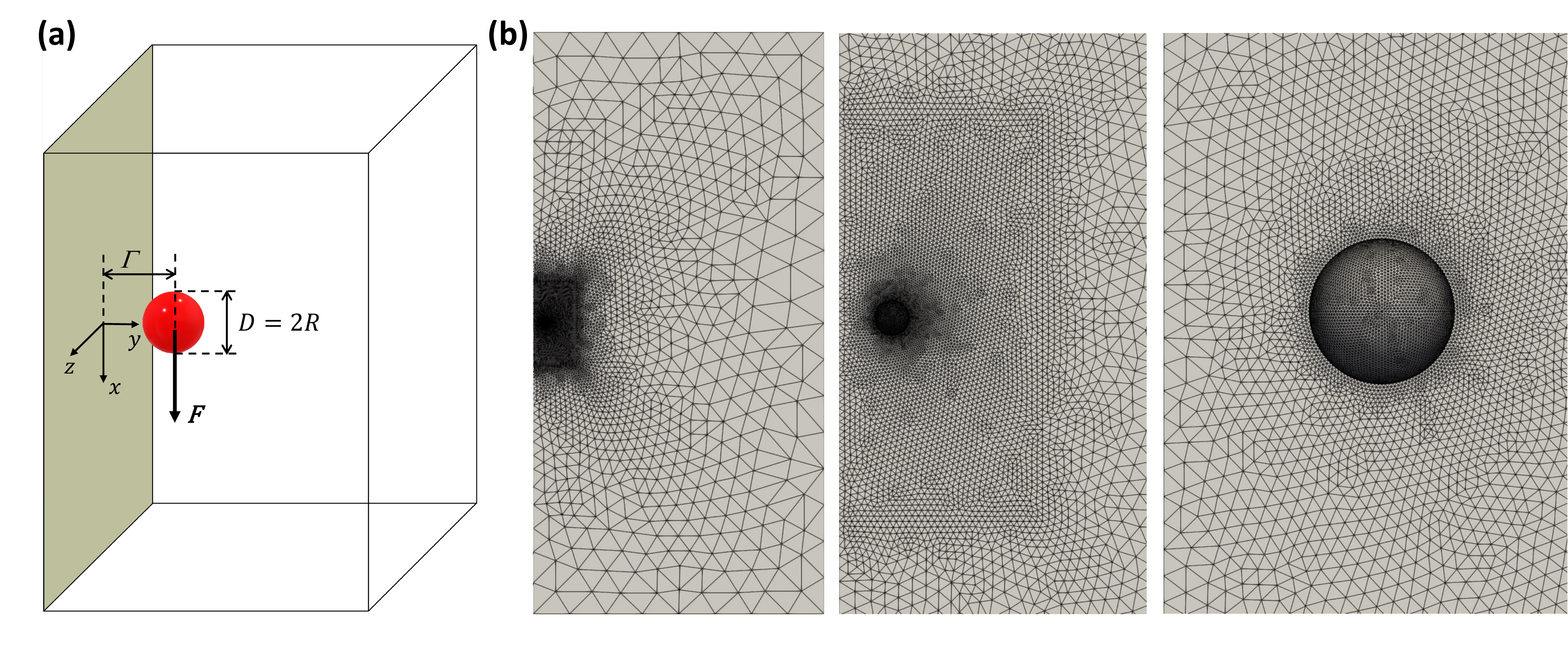}
\caption{(a) Schematic representation of the computational domain: a spherical particle with diameter $D$ subjected to a constant force settles in an EVP fluid near a wall; the origin of a Cartesian reference frame is set on the wall with $x$ the direction of the applied force; the distance between the particle center and the wall is denoted by $\Gamma$. (b) Typical mesh used in the simulations. The panels show the mesh over the particle surface and on the symmetry plane with the particle progressively zoomed from the left to the right.} 
\label{fig:scheme}
\end{figure}

The dynamics of a spherical particle settling in an EVP fluid near a wall is simulated in the domain shown in Fig.~\ref{fig:scheme}a. A Cartesian reference frame is selected with origin at the center of the physical wall with axes oriented as in figure. A force acting along the $x-$direction is applied on the spherical particle with radius $R$. The distance between the particle center and the wall is denoted by $\Gamma$. The domain size is chosen sufficiently large in order to avoid any influence of the box boundaries (except, of course, the boundary representing the physical wall) on the particle motion. From preliminary tests by progressively increasing the domain size, we found that a box with side lengths of $160R\times80R\times160R$ in the $x$, $y$ and $z$ directions, respectively, fulfills the previous condition for any set of parameters investigated in this work.

Assuming inertialess flow conditions, the motion of the fluid domain is governed by the continuity and momentum balance equations:
\begin{eqnarray}
	\nabla\cdot \boldsymbol{u}=0
	\label{eqn:continuity}
\end{eqnarray}
\begin{eqnarray}
	\nabla\cdot\boldsymbol{T}=\boldsymbol{0}
	\label{eqn:momentum}
\end{eqnarray}
where $\boldsymbol{u}$ is the fluid velocity and $\boldsymbol{T}$ is the total stress tensor that can be written as:
\begin{eqnarray}
	\boldsymbol{T}=-p\boldsymbol{I}+2\eta_\mathrm{s}\boldsymbol{D}+\boldsymbol{\tau} \label{eqn:viscoelastic constitutive}
\end{eqnarray}
In this equation, $p$ is the pressure, $\boldsymbol{I}$ is the unity tensor, $\eta_\mathrm{s}$ is the viscosity of a Newtonian solvent, $\boldsymbol{D}=(\nabla\boldsymbol{u}+(\nabla\boldsymbol{u})^\mathrm{T})/2$ denotes the rate-of-deformation tensor, and $\boldsymbol{\tau}$ is the extra stress due to the non-Newtonian nature of the fluid. In this work, we model the EVP fluid with the Giesekus constitutive equation modified as proposed by Saramito \cite{Saramito2007new} to account for the yield stress:
\begin{eqnarray}
	\lambda\stackrel{\nabla}{\boldsymbol{\tau}}+\max\left(0,\frac{|\boldsymbol{\tau}_\text{d}|-\tau_\text{y}}{|\boldsymbol{\tau}_\text{d}|}\right)\left(\boldsymbol{\tau}+\frac{\alpha\lambda}{\eta_\mathrm{p}}\boldsymbol{\tau}\cdot\boldsymbol{\tau}\right)=2\eta_\mathrm{p}\boldsymbol{D}
	\label{eqn:Saramito_GSK}
\end{eqnarray}
where $|\boldsymbol{\tau}_\text{d}|$ is given by:
\begin{equation}
	|\boldsymbol{\tau}_\text{d}|=\sqrt{\frac{1}{2}\hat{\boldsymbol{\tau}}:\hat{\boldsymbol{\tau}}} \label{eqn:magn_tau}
\end{equation}
with:
\begin{eqnarray}
	\hat{\boldsymbol{\tau}}=\boldsymbol{\tau}-\frac{1}{3}\text{Tr}\boldsymbol{\tau}\,\boldsymbol{I}
	\label{eqn:tau_hat}
\end{eqnarray}
with `Tr' the trace operator. In Eq.~\eqref{eqn:Saramito_GSK}, $\lambda$ is the relaxation time,  $\eta_\text{p}$ is the polymer contribution to the viscosity, $\alpha$ is the `mobility' parameter, and $\tau_\text{y}$ is the yield stress. The symbol $(^{\nabla})$ denotes the upper-convected time derivative, defined as:
\begin{eqnarray}
	\stackrel{\nabla}{\boldsymbol{\tau}}\equiv\frac{\partial\boldsymbol{\tau}}{\partial t}+\boldsymbol{u}\cdot
	\nabla\boldsymbol{\tau}-(\nabla\boldsymbol{u})^{T}\cdot\boldsymbol{\tau}-\boldsymbol{\tau}\cdot
	\nabla\boldsymbol{u} \label{eqn:upper-convected}
\end{eqnarray}

The rheological properties of the Saramito-Giesekus model in simple shear and uniaxial extensional flows are reported in the Appendix.

As previously mentioned, a force $\boldsymbol{F}$ is applied on the particle along the $x-$direction. Assuming inertialess conditions for the particle, the following equations hold at the particle surface $S$:
\begin{eqnarray}
	\boldsymbol{F}=(F,0,0)=\int_S\boldsymbol{T}\cdot \boldsymbol{n}\,dS\label{eqn:force}
\end{eqnarray}
\begin{eqnarray}
	\int_S(\boldsymbol{x}-\boldsymbol{X})\times(\boldsymbol{T}\cdot\boldsymbol{n})\,dS=\boldsymbol{0} \label{eqn:torque}
\end{eqnarray}
where $\boldsymbol{X}$ is the particle centroid, $\boldsymbol{x}$ is the position vector of a point on the boundary $S$, $\boldsymbol{n}$ is the outwardly directed unit normal vector on $S$ and $dS$ is the local surface area.

Since the force is applied along the $x-$direction and the system is unconfined along the $z-$direction, the $xy-$plane (passing through the particle center) is a symmetry plane. Thus, we can consider a computational domain that is one-half of the full domain. Regarding the boundary conditions, adherence is assumed at the particle boundary, resulting in the rigid-body motion equation:
\begin{equation}
	\boldsymbol{u}=\boldsymbol{U}+\boldsymbol{\omega}\times(\boldsymbol{x}-\boldsymbol{X}) \label{eqn:rigid-body}
\end{equation}
where $\boldsymbol{U}$ and $\boldsymbol{\omega}$ are the translational and angular velocities of the particle. At the external boundaries of the computational domain, no-slip conditions are imposed. The symmetry boundary condition is applied at the symmetry plane. Due to the inertialess assumption, no initial condition for the velocity is needed. The stress-free condition, i.e., $\boldsymbol{\tau}|_{t=0}=\boldsymbol{0}$, is considered as initial condition for the viscoelastic stress.

Once the fluid velocity, pressure and stress fields are calculated along with the particle kinematic quantities, the particle center $\boldsymbol{X}$ and angle $\boldsymbol{\Theta}$ are updated by integrating the following equations:
\begin{eqnarray}
	\frac{d\boldsymbol{X}}{dt}=\boldsymbol{U},~~~\boldsymbol{X}|_{t=0}=\boldsymbol{X}_{\mathrm{0}}
	\label{eqn:kinematic X}
\end{eqnarray}
\begin{eqnarray}
	\frac{d\boldsymbol{\Theta}}{dt}=\boldsymbol{\omega},~~~\boldsymbol{\Theta}|_{t=0}=
	\boldsymbol{\Theta}_{\mathrm{0}}
	\label{eqn:kinematic Theta}
\end{eqnarray}
where $\boldsymbol{X}_{\mathrm{0}}$ and $\boldsymbol{\Theta}_{\mathrm{0}}$ are the initial particle position and orientation. For a sphere, the latter equation is not relevant.

\subsection{Dimensionless equations}
\label{sec:Dimensionless equations}

It is useful to present the set of equations discussed above in dimensionless form. The particle radius $R$ is chosen as characteristic length. Following Fraggedakis et al. \cite{Fraggedakis2016yielding} and Sarabian et al. \cite{Sarabian2022interface}, we selected the characteristic velocity as the one obtained by balancing viscous and
buoyancy forces $U_\text{c}=3F/(4\pi R\,\eta_0)$. Consequently, the characteristic time is $t_\text{c}=R/U_\text{c}$. The characteristic stress is $\tau_\text{c}=\eta_0/t_\text{c}$ where $\eta_0=\eta_\text{s}+\eta_\text{p}$ is the zero-shear viscosity.

Denoting with starred symbols the dimensionless quantities, Eqs.~\eqref{eqn:continuity}-\eqref{eqn:viscoelastic constitutive} can be expressed in dimensionless form as:
\begin{eqnarray}
	\nabla^{*}\cdot\boldsymbol{u}^{*}=0 \label{eqn:continuity_dimensionless}
\end{eqnarray}
\begin{eqnarray}
	-\nabla^{*}p^{*}+\eta_\mathrm{r}\nabla^{*2} \boldsymbol{u}^{*}+\nabla^{*}\cdot\boldsymbol{\tau}^{*}=\boldsymbol{0}\label{eqn:momentum_dimensionless}
\end{eqnarray}
Notice that the expression of the total stress tensor in Eq.~\eqref{eqn:viscoelastic constitutive} has been substituted in the momentum balance. The dimensionless constitutive equation is:
\begin{eqnarray}
	Wi\stackrel{\nabla}{\boldsymbol{\tau}}{}^{*}+\max\left(0,\frac{|\boldsymbol{\tau}^*_\text{d}|-Bn}{|\boldsymbol{\tau}^*_\text{d}|}\right)\left(\boldsymbol{\tau}^{*}+\frac{\alpha Wi}{1-\eta_\text{r}}\,\boldsymbol{\tau}^{*}\cdot\boldsymbol{\tau}^{*}\right)=2(1-\eta_\mathrm{r})\boldsymbol{D}^{*}
	\label{eqn:GSK model_dimensionless}
\end{eqnarray}

In the above equations, the Weissenberg and Bingham numbers, and the viscosity ratio appear: 
\begin{equation}
	Wi=\frac{\lambda}{t_\mathrm{c}}=\frac{\lambda F}{\frac{4}{3}\pi R^2\eta_0} \label{eqn:Wi_number}
\end{equation}
\begin{equation}
	Bn=\frac{\tau_\text{y} R}{\eta_0 U_\text{c}}=\frac{\tau_\text{y}\frac{4}{3}\pi R^2}{F} \label{eqn:Bn_number}
\end{equation}
\begin{eqnarray}
	\eta_\mathrm{r}=\frac{\eta_\mathrm{s}}{\eta_\mathrm{0}} \label{eqn:visc_number}
\end{eqnarray}
Along with these three dimensionless parameters, we also have the mobility $\alpha$ and the dimensionless particle-wall distance $\Gamma^*=\Gamma/R$.

In this work, we fix $\alpha=0.2$ (denoting a shear-thinning fluid) and $\eta_\text{r}=0.1$. The particle dynamics will be studied by varying $Wi$, $Bn$, and $\Gamma^*$. From now on, we remove the starred symbols and all the quantities are intended to be dimensionless.

\subsection{Numerical method, convergence, and validation}
\label{sec:Numerical method, convergence, and validation}

The governing equations are solved by the finite element method with the DEVSS-G/SUPG formulation \cite{Guenette1995new,Bogaerds2002stability,Brooks1982streamline} and the log-representation for the conformation tensor \cite{Fattal2004constitutive,Hulsen2005flow} to improve the convergence at high Weissenberg numbers. The rigid-body motion is imposed through constraints in each node of the particle surface by means of Lagrange multipliers \cite{DAvino2008rotation}. The Arbitrary Lagrangian-Eulerian (ALE) moving mesh method is adopted to handle the particle motion \cite{Hu2001direct}. As previously done \cite{DAvino2010viscoelasticity,Villone2011numerical}, the translational velocity of the particle along the sedimentation direction is subtracted from the convective term of the constitutive equation so that the particle center of volume is fixed along $x$ and only moves along $y$ due to the migration. In this way, no remeshing is needed during a simulation.

The code for a viscoelastic suspending fluid has been extensively validated in numerous previous publications (see, e.g., Refs. \cite{DAvino2010viscoelasticity,Villone2011numerical,DAvino2011migration}). The Saramito-Giesekus constitutive equation employed in this work is a modified Giesekus model where a factor (that contains the viscoelastic stress components through $\boldsymbol{\tau}_\text{d}$) multiplies the term $(\boldsymbol{\tau}+\alpha\lambda/\eta_\text{p}\boldsymbol{\tau}\cdot\boldsymbol{\tau})$. In our implementation, such a term is treated explicitly \cite{DAvino2010numerical}, i.e., the viscoelastic stress components are taken from the previous time step. Hence, the Saramito factor is straightforwardly implemented in the discretization of the constitutive equation.

\begin{figure}[t!]
\centering
\includegraphics[width=0.5\textwidth]{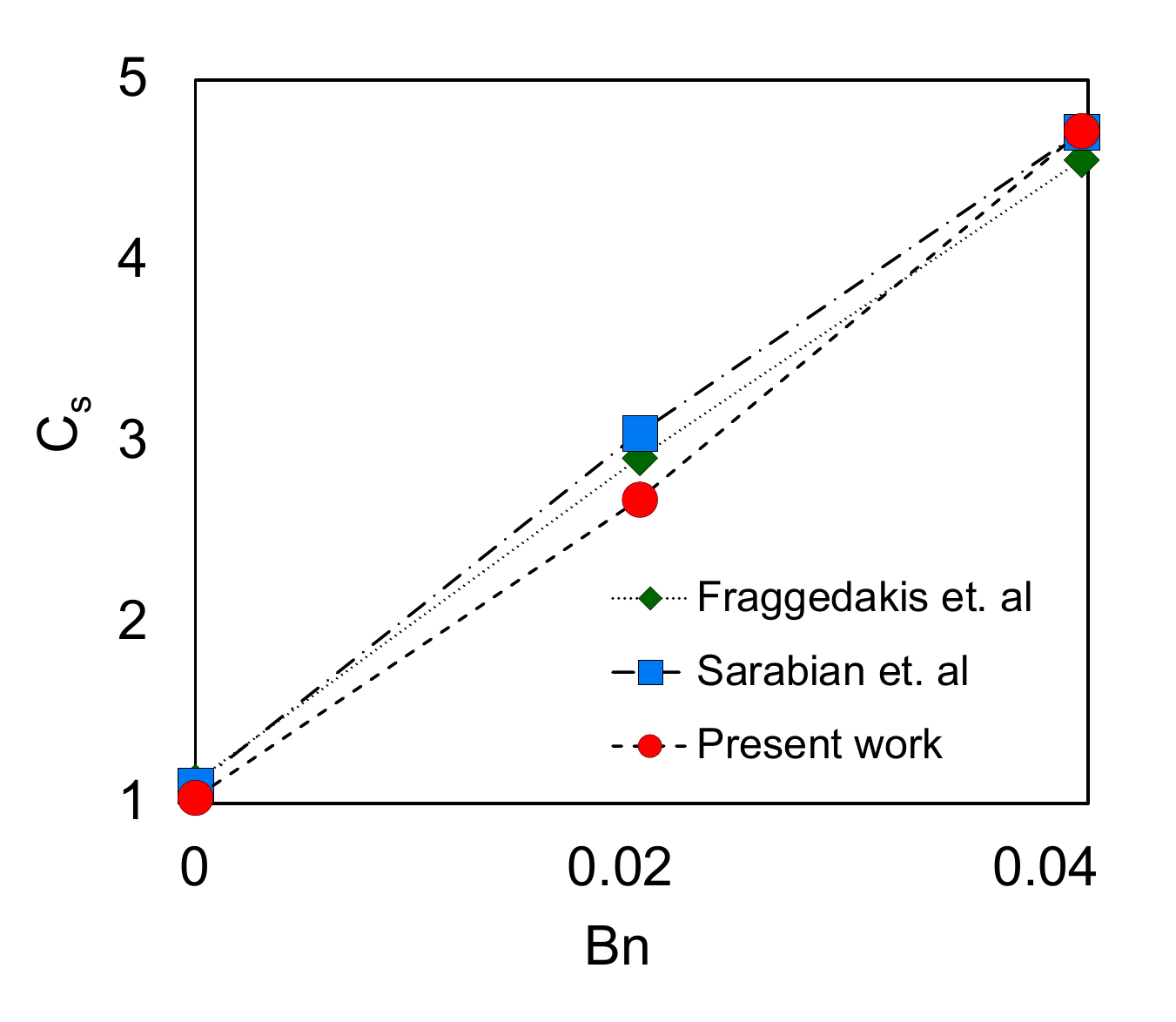}
\caption{Stokes drag coefficient of a spherical particle sedimenting in an unconfined EVP fluid as a function of the Bingham number for $Wi=1$ computed in Fraggedakis et al. \cite{Fraggedakis2016yielding}, Sarabian et al. \cite{Sarabian2022interface}, and the present work. The fluid is modelled by the Saramito constitutive equation obtained by setting $\alpha=0$ in Eq.~\eqref{eqn:Saramito_GSK}.}
\label{fig:validation}
\end{figure}

We have first validated the code without particle by solving the transient flow field of an EVP fluid in a 2D periodic domain with an applied shear flow. The original Saramito model (obtained by setting $\alpha=0$ in Eq.~\eqref{eqn:Saramito_GSK}) is employed. The results (not reported) show an excellent agreement between the analytical and numerical solution for all the investigated combinations of $Wi$ and $Bn$.

The code is further validated by including the particle. The case of a sedimenting sphere in an EVP fluid modelled through the Saramito constitutive equation in an infinite medium is considered. The computed steady-state settling velocity $U_\text{s}$ is used to evaluated the Stokes drag coefficient defined as:
\begin{equation}
C_\text{S}=\frac{F}{6\pi\eta_0 U_\text{s}R}
\label{eqn:Stokes_drag}
\end{equation}
Figure~\ref{fig:validation} compares the numerical results between the present work, Fraggedakis et al. \cite{Fraggedakis2016yielding} and Sarabian et al. \cite{Sarabian2022interface} for $Wi=1$ and three values of the Bingham number. The agreement is quantitatively good and the small deviations may be attributed to the different numerical methods employed in these works.

\begin{figure}[t!]
\centering
\includegraphics[width=1\textwidth]{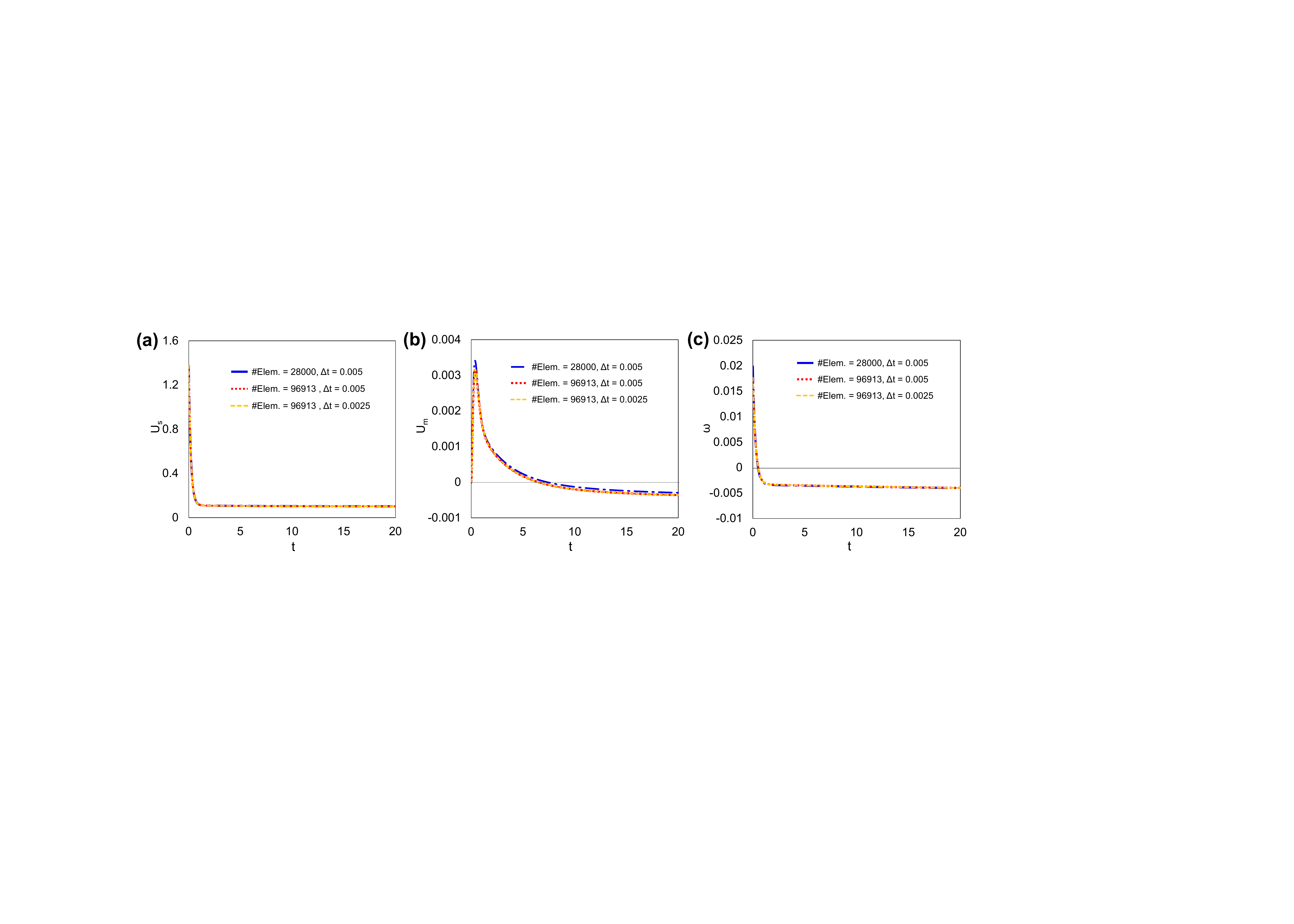}
\caption{Sedimentation (a), migration (b), and angular (c) velocities of a spherical particle in an EVP fluid near a wall as a function of time for different mesh resolutions and time step sizes. The parameters are: $Wi=2$, $Bn=0.01$, $\Gamma=1.5$.}
\label{fig:convergence}
\end{figure}

Finally, mesh and time convergence are verified for all the simulations. Figure~\ref{fig:scheme}b displays a typical mesh employed in this work. The images show the mesh over the particle surface and on the symmetry plane, with the particle progressively zoomed from the left to the right. Notice that the particle is `surrounded' by a box (clearly visible in the middle image of Fig~\ref{fig:scheme}b) with a finer resolution. We need such a refinement in order to accurately capture the yielded/unyielded regions as discussed later. Typical results of convergence tests are reported in Fig.~\ref{fig:convergence} where the sedimentation $U_\text{s}$, migration $U_\text{m}$, and angular velocity $\omega$ are reported as a function of time for two mesh resolutions and two time step sizes. The parameters are: $Wi=2$, $Bn=0.01$, $\Gamma=1.5$. For all the kinematic quantities, the curves fairly overlap. As expected, the required mesh resolution and time step size to get convergent results depend on $Wi$ and $Bn$. For $Wi$ up to 2 and $Bn$ up to 0.01, about 45000 tetrahedral elements are needed in the domain, with about 1000 triangular elements over the particle surface. A (dimensionless) time step size of 0.005 is used. At higher values of $Wi$ and $Bn$, the number of elements need to be increased up to 75000, with about 1600 triangular elements over the particle, and the time step size reduced to 0.0025.

\section{Results}
\label{sec:Results}

\begin{figure}[t!]
\centering
\includegraphics[width=1\textwidth]{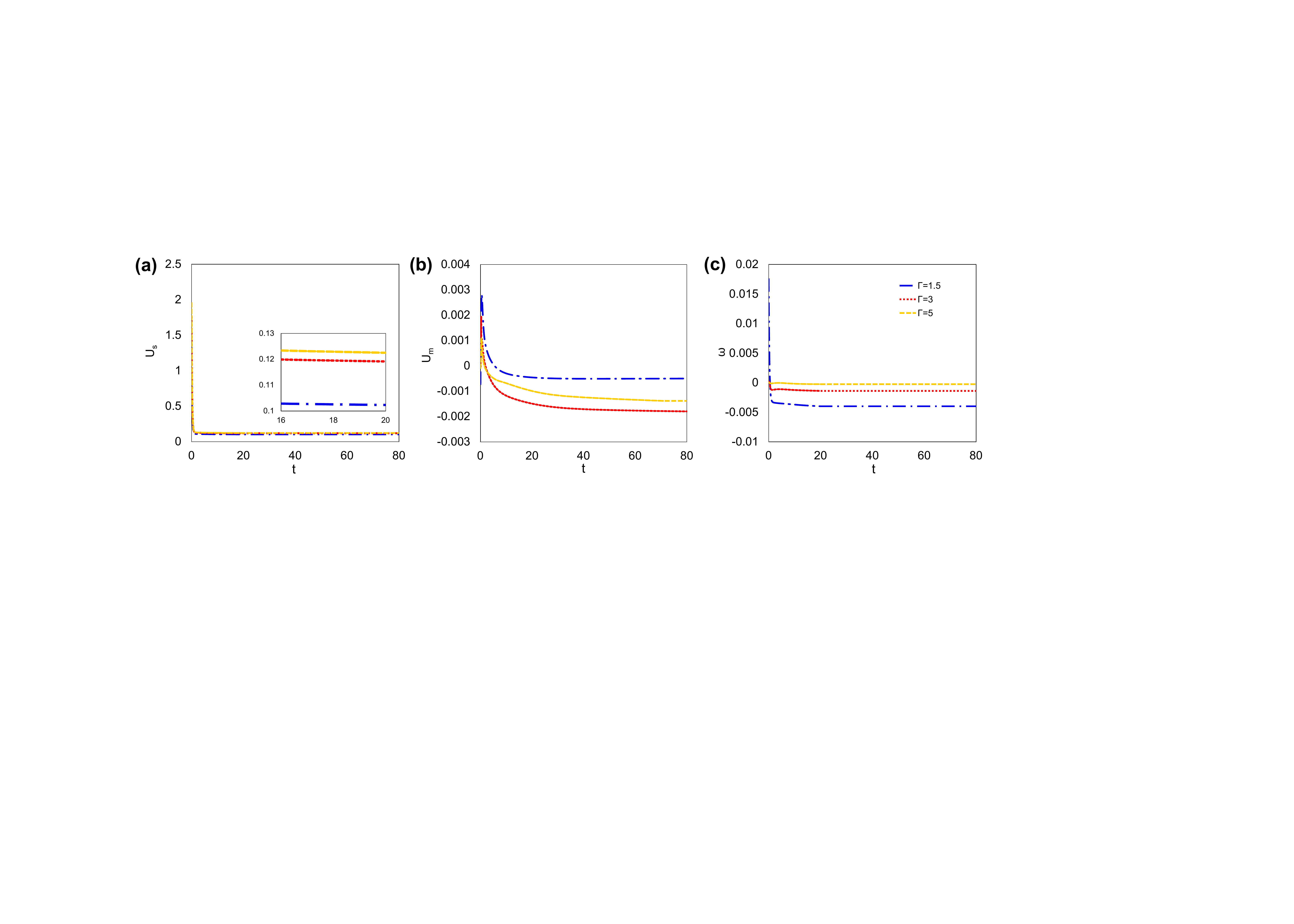}
\caption{Sedimentation (a), migration (b), and angular (c) velocities of a spherical particle in an EVP fluid as a function of time for different particle-wall distances. The Weissenberg number is $Wi=2$ and the Bingham number is $Bn=0.01$ The inset in panel (a) shows a zoom of the three curves at long times once the trends are nearly horizontal.} 
\label{fig:transient}
\end{figure}

The particle dynamics in terms of the translational and rotational velocities is first addressed, with a special focus on the migration phenomenon as the particle-wall distance, Weissenberg and Bingham numbers are varied. The flow field around the particle and the yielded/unyielded regions are then presented. The purely viscoelastic case is also simulated to understand the effect of plasticity on the particle motion.

The sedimentation velocity of the particle as a function of time is displayed in Fig.~\ref{fig:transient}a for $Wi=2$ and $Bn=0.01$, and for three values of the particle-wall distance, $\Gamma=1.5$, $\Gamma=3$, and $\Gamma=5$. At the initial time, the viscoelastic stresses are zero so the system is equivalent to a Stokes flow. The first point on the curves is, in fact, the settling velocity of a sphere in a Newtonian liquid with viscosity $\eta_\text{s}$ at a certain distance from a flat wall. Since we set $\eta_\text{s}/\eta_\text{0}=0.1$, the Newtonian sedimentation velocity is one order of magnitude higher than the one in a viscoelastic fluid after the stress development (where the characteristic viscosity is $\eta_0$). Indeed, after the initial transient, the settling velocity rapidly reduces and the trend becomes nearly horizontal. As expected, the long-time value of the sedimentation rate depends on the particle-wall distance as visible in the inset of the figure. It has to be pointed out, however, that the value reached at long times is not constant as the particle-wall distance is changing due to the migration phenomenon (see below). This behavior is similar to what observed for the translational velocity of a particle suspended in a viscoelastic fluid under planar shear \cite{DAvino2010viscoelasticity} or pressure-driven flow \cite{Villone2011numerical}. In the present case, indeed, we find that the curves taken after the initial stress development at different particle-wall distances superimpose and identify a mastercurve. As thoroughly discussed in previous works \cite{DAvino2010viscoelasticity,DAvino2010numerical,Villone2011numerical}, the existence of a mastercurve suggests that the settling velocity (but this holds for the other kinematic quantities as well, see below) is uniquely related to the particle-wall distance once the viscoelastic stresses fully develop.

\begin{figure}[t!]
\centering
\includegraphics[width=1\textwidth]{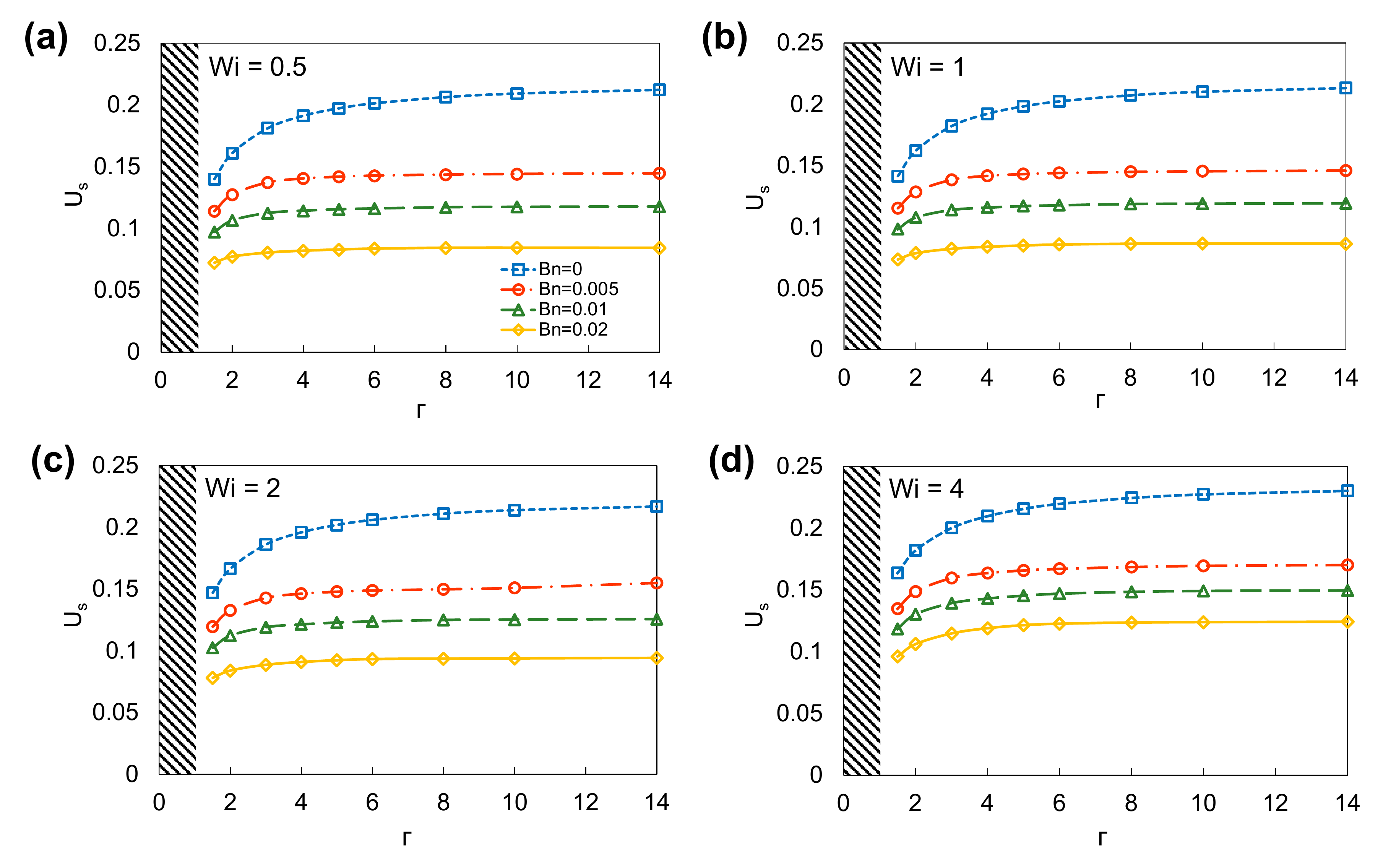}
\caption{Sedimentation velocity of the particle as a function of the particle-wall distance for different values of the Weissenberg and Bingham number. The shaded area is the region inaccessible to the particle due to its finite size.} 
\label{fig:Us}
\end{figure}

Figure~\ref{fig:Us} shows the mastercurves of the particle sedimentation rate for four values of the Weissenberg number. In each plot, four values of the Bingham number are considered. The case at $Bn=0$ corresponds to the purely viscoelastic case. It can be readily observed that: i) the presence of a confining wall hinders the sedimentation speed, as also found for a Newtonian fluid \cite{Happel1983low}, ii) higher Bingham numbers slow down the sedimentation, in agreement with unconfined systems \cite{Fraggedakis2016yielding,Sarabian2022interface}, iii) at a particle-wall distance of about 10 radii, the curves become nearly horizontal denoting that the wall has a negligible effect in the settling phenomenon, i.e., the system can be considered unconfined, iv) increasing the Weissenberg number speeds up the sedimentation although the effect is quantitatively small. Regarding the effect of the Weissenberg number on the sedimentation speed, a vast literature exists for a spherical particle in a viscoelastic fluid (see, e.g., Ref.~\cite{Mckinley2002transport}). Previous simulation results in a variety of fluids show that the fluid viscoelasticity reduces the drag coefficient as $Wi$ increases, i.e., the sphere settles faster in a viscoelastic fluid, in agreement with the data shown in Fig.~\ref{fig:Us}. In these works, the characteristic flow time used to define the Weissenberg number is the particle radius divided by the steady-state sedimentation speed. By using this definition, the range of $Wi$ corresponding to the purely viscoelastic case in an unbounded domain (rightmost black point in the panels of Fig.~\ref{fig:Us}) is between 0.1 and 0.9. In this range, the effect of $Wi$ on the sedimentation speed is relatively weak \cite{Faroughi2020closure}, as confirmed by our simulation results.

\begin{figure}[t!]
\centering
\includegraphics[width=1\textwidth]{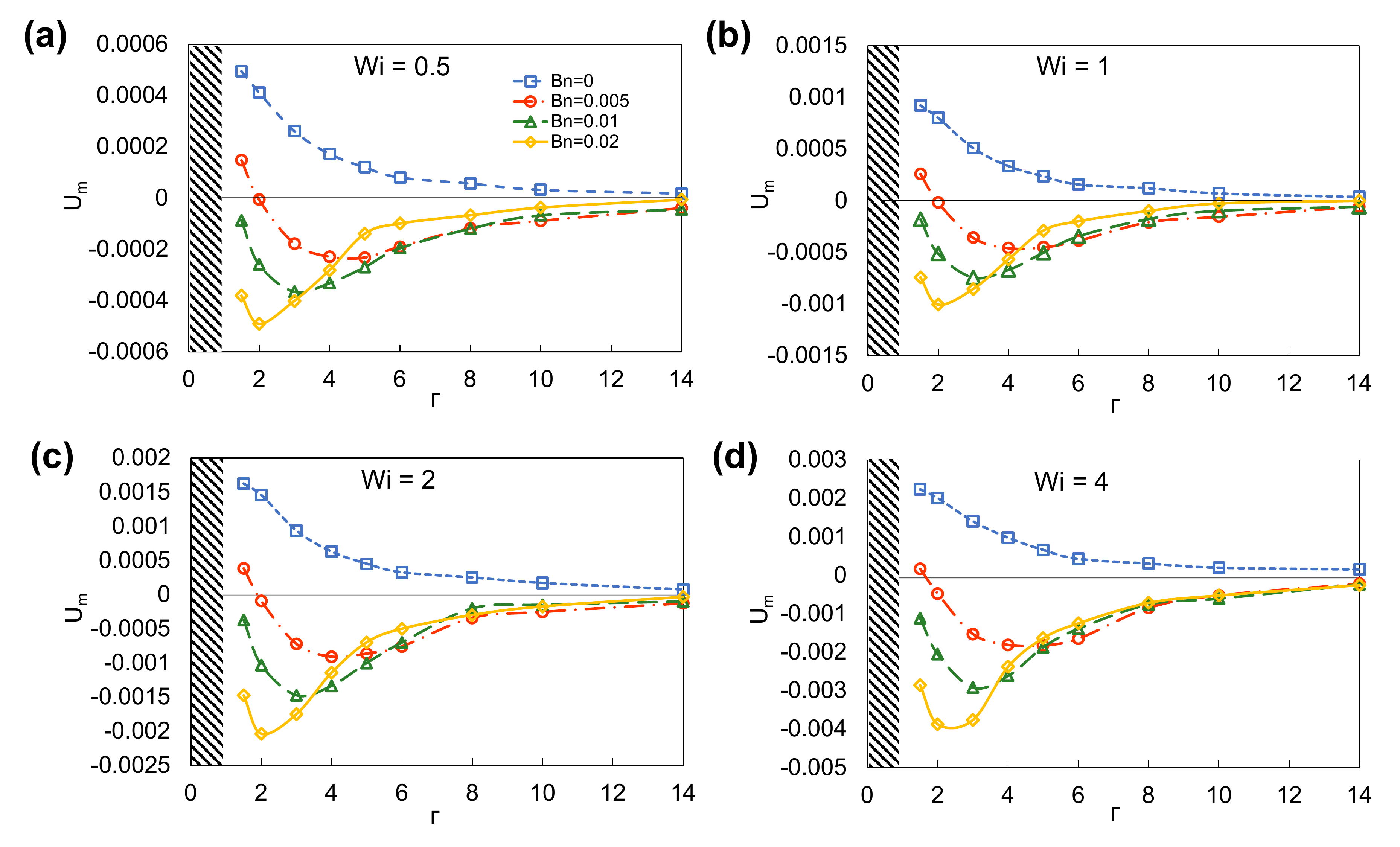}
\caption{Migration velocity of the particle as a function of the particle-wall distance for different values of the Weissenberg and Bingham number. The shaded area is the region inaccessible to the particle due to its finite size.} 
\label{fig:Um}
\end{figure}

As mentioned in the Introduction, a particle immersed in a viscoelastic fluid near a wall and subjected to a force parallel to the wall experiences a lateral motion termed `migration' \cite{Singh2000sedimentation,Spanjaards2019numerical}. Figure~\ref{fig:transient}b reports the transient migration velocity (denoted as $U_\text{m}$) for $Wi=2$, $Bn=0.01$ and three distances from the wall. The three curves start at the origin as for a sphere in a Newtonian inertialess fluid no migration occurs. As soon as the viscoelastic stresses start develop, a maximum is achieved after which the curves attain a nearly constant trend. Similarly to the sedimentation velocity previously discussed, the curves taken after the initial stress development at different particle-wall distances superimpose describing a mastercurve. Figure~\ref{fig:Um} shows the mastercurves of the migration velocity at different values of the Weissenberg and Bingham numbers. Positive or negative values of the migration velocity indicate a lateral motion of the particle away or towards the wall, respectively. In a purely viscoelastic case ($Bn=0$), the migration velocity is positive for any particle-wall distance. The trend is monotonic and decreasing as the particle moves away from the wall. At the largest investigated distance ($\Gamma=14$), the migration velocity is nearly zero denoting a negligible effect of the confinement (in an unbounded domain a spherical particle does not migrate). The curves at varying $Wi$ are qualitatively similar. These results agree with those reported by Spanjaards et al. \cite{Spanjaards2019numerical} for a Couette flow cell. In this regard, it should be remarked that the simulation results reported in the aforementioned paper show that the migration velocity is positive (wall repulsion) for $Wi$ below a threshold and becomes negative (wall attraction) for large $Wi$-values. The maximum Weissenberg number explored in the present work is lower than the threshold reported in Spanjaards et al. \cite{Spanjaards2019numerical}, confirming the wall repulsive behavior. (Notice that the definition of the Weissenberg number in Spanjaards et al. \cite{Spanjaards2019numerical} is different from the one adopted in this paper; specifically, for the same set of parameters, our $Wi$ is nine times lower than the one given in Spanjaards et al. \cite{Spanjaards2019numerical}.) Higher values of the Weissenberg number lead to a faster migration rate as previously found for different flow fields \cite{DAvino2010viscoelasticity,Villone2011numerical, DAvino2010numerical}.

As the Bingham number increases, the migration velocity progressively decreases and changes sign, i.e., the particle inverts its migration direction and moves towards the wall. The trend becomes non-monotonic with the migration velocity that achieves a minimum as the particle-wall distance increases, and approaches to zero for large distances. Interestingly, for $Bn=0.005$, the migration velocity crosses the horizontal axis at a particle-wall distance of about $\Gamma=2$. This is a stable equilibrium point, i.e., the particle settles maintaining a fixed distance from the wall. A similar behavior could occur for $Bn=0.01$ too since the extrapolated curves likely cross the horizontal axis at a finite (although very small) particle-wall distance. For higher values of $Bn$, the migration velocity becomes negative for any $\Gamma$-value thus the particle sediments near the wall (in theory, in contact with the wall). In this regard, we mention that the simulations break down for very low particle-wall distances so we were unable to investigate the particle dynamics close to the wall (the minimum initial distance simulated is $\Gamma=1.5$ corresponding to one-half radius between the wall and the particle surface). A variation of the Weissenberg number speeds up the migration velocity also for an elastoviscoplastic fluid, although the trends of the migration velocity curves are essentially unaffected. The only visible difference is that the curve at $Bn=0.005$ for $Wi=4$ crosses the $x$-axis at about $\Gamma=1.5$, thus the wall repulsive region is narrower as compared to lower $Wi$-values. Higher values of the Weissenberg number are expected to further reduce this region and lead to a fully repulsive dynamics from the wall for smaller values of the Bingham number.

\begin{figure}[t!]
\centering
\includegraphics[width=1\textwidth]{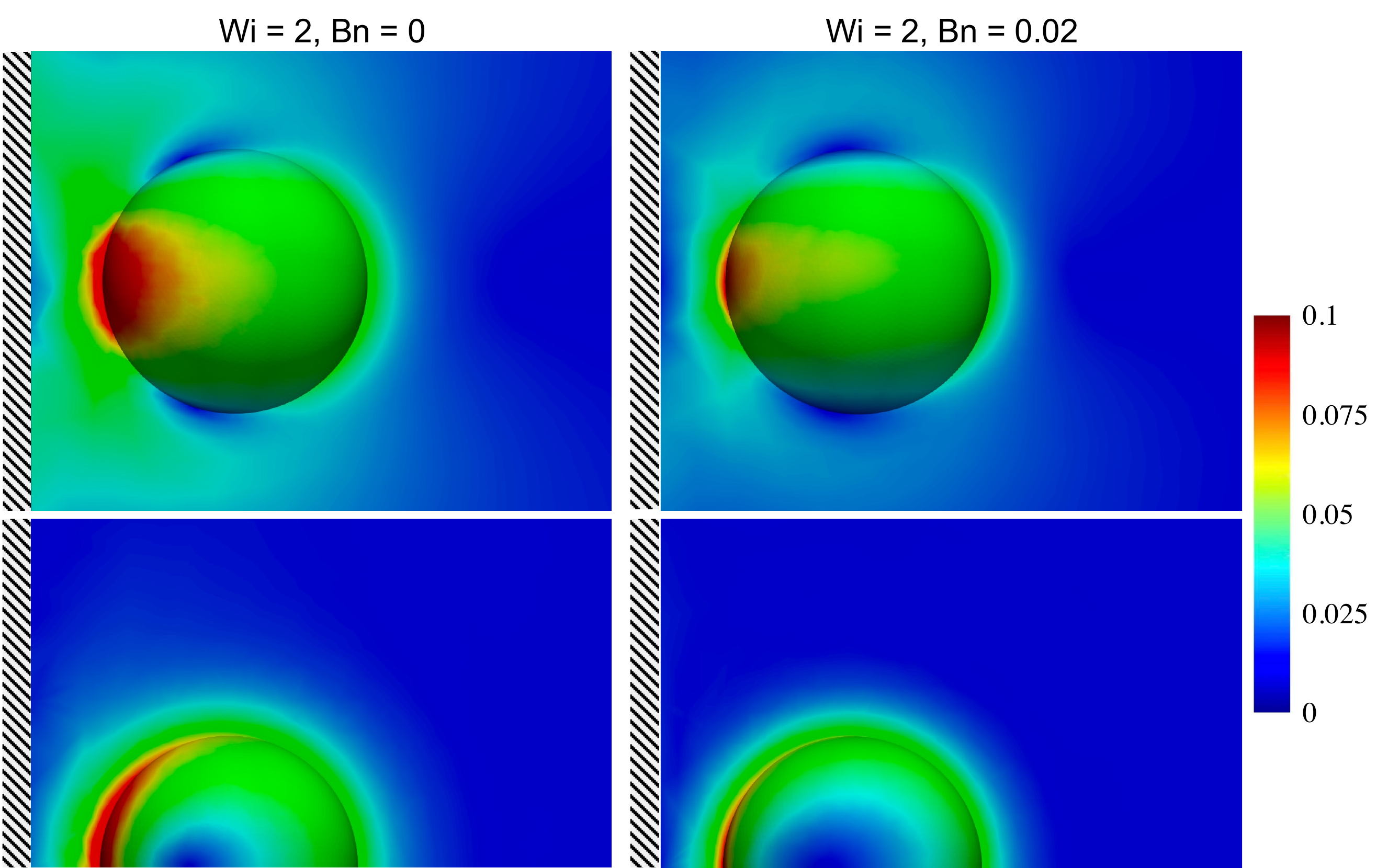}
\caption{Shear rate distribution on the $xy-$plane and on the spherical surface (top), and on the $yz-$plane (bottom). The purely viscoelastic case ($Bn=0$) is considered on the left. On the right the EVP case is shown at $Bn=0.02$. In both cases, the Weissenberg number is $Wi=2$. The shaded area is the region inaccessible to the particle due to its finite size.} 
\label{fig:shearrate}
\end{figure}

The results just discussed show that fluid plasticity has a relevant effect on the migration phenomenon as the particle lateral motion reverts direction. Previous works on the dynamics of particles on viscoelastic fluids in shear and Poiseuille flows attributed the migration mechanism to an imbalance of normal stresses due to the shear rate gradients around the particle that generates a net force displacing the particle trajectory \cite{DAvino2017particle}. The shear rate distribution around the particle, indeed, can provide useful information about the migration dynamics. The top panels of Fig.~\ref{fig:shearrate} display the distribution of the (dimensionless) shear rate $\dot{\gamma}=\sqrt{2\boldsymbol{D}:\boldsymbol{D}}$ on the symmetry $xy-$plane and on the spherical surface. The bottom panels report the same quantity on the $yz$-plane. On the left, the purely viscoelastic case ($Bn=0$) is considered at Weissenberg number $Wi=2$. The right panels refer to the elastoviscoplastic fluid with $Bn=0.02$ and the same value of $Wi$ as in the left figures. In both cases, the distance from the channel inlet is $\Gamma=1.5$. The first case corresponds to the leftmost data point of the blue curve in Fig.~\ref{fig:Um}c, showing a positive value of the migration velocity, i.e., the particle is moving away from the wall. On the contrary, the migration velocity for the EVP fluid case is represented by the leftmost data point of the yellow curve in Fig.~\ref{fig:Um}c which is negative, i.e., the particle is moving towards the wall.

As expected, the highest shear rate region is found in between the particle and the wall where the gap is small. Specifically, the shear rate is higher for the viscoelastic case (see the large red region on the left side of the particle surface). This is a consequence of the higher sedimentation speed of the particle in a viscoelastic fluid rather than in an elastoviscoplastic one, as previously discussed (see Fig.~\ref{fig:Us}). Nevertheless, the migration velocity has a similar magnitude in the two cases (around 0.0015, as shown in Fig.~\ref{fig:Um}c), although opposite in sign. Recalling that the migration phenomenon is related to the shear rate gradient around the particle, Fig.~\ref{fig:shearrate} shows that the spatial variation of the shear rate on the left side of the particle from the surface to the wall is larger for the viscoelastic fluid as compared to the EVP one. This generates, in turn, a stronger net force away from the wall in the viscoelastic case. On the contrary, the shear rate gradient on the right side of the particle is higher for the EVP case. Indeed, the shear rate value at the particle surface is the same for the two cases (denoted by a green color corresponding to about 0.05). However, it becomes near zero (blue color) in a shorter distance from the sphere in the EVP case, as visible from the narrower green band at the right side of the particle in the right panels of Fig.~\ref{fig:shearrate}. This generates a force towards the wall that is stronger for the EVP fluid. Such a force overcomes the (small) force from the left side, pushing the particle towards the wall. The larger shear rate gradient on the right side observed in the EVP case is likely induced by the presence of the unyielded region that `confines' the sphere on the right (see below). As the Bingham number increases, the yielded region moves closer to the particle, further increasing the shear rate gradient on the unconfined side, leading to a faster migration towards the wall. By further increasing the Bingham number, we expect that the unyielded region touches the particle leading to the blockage (no sedimentation \cite{Fraggedakis2016yielding} and no migration).

\begin{figure}[t!]
\centering
\includegraphics[width=1\textwidth]{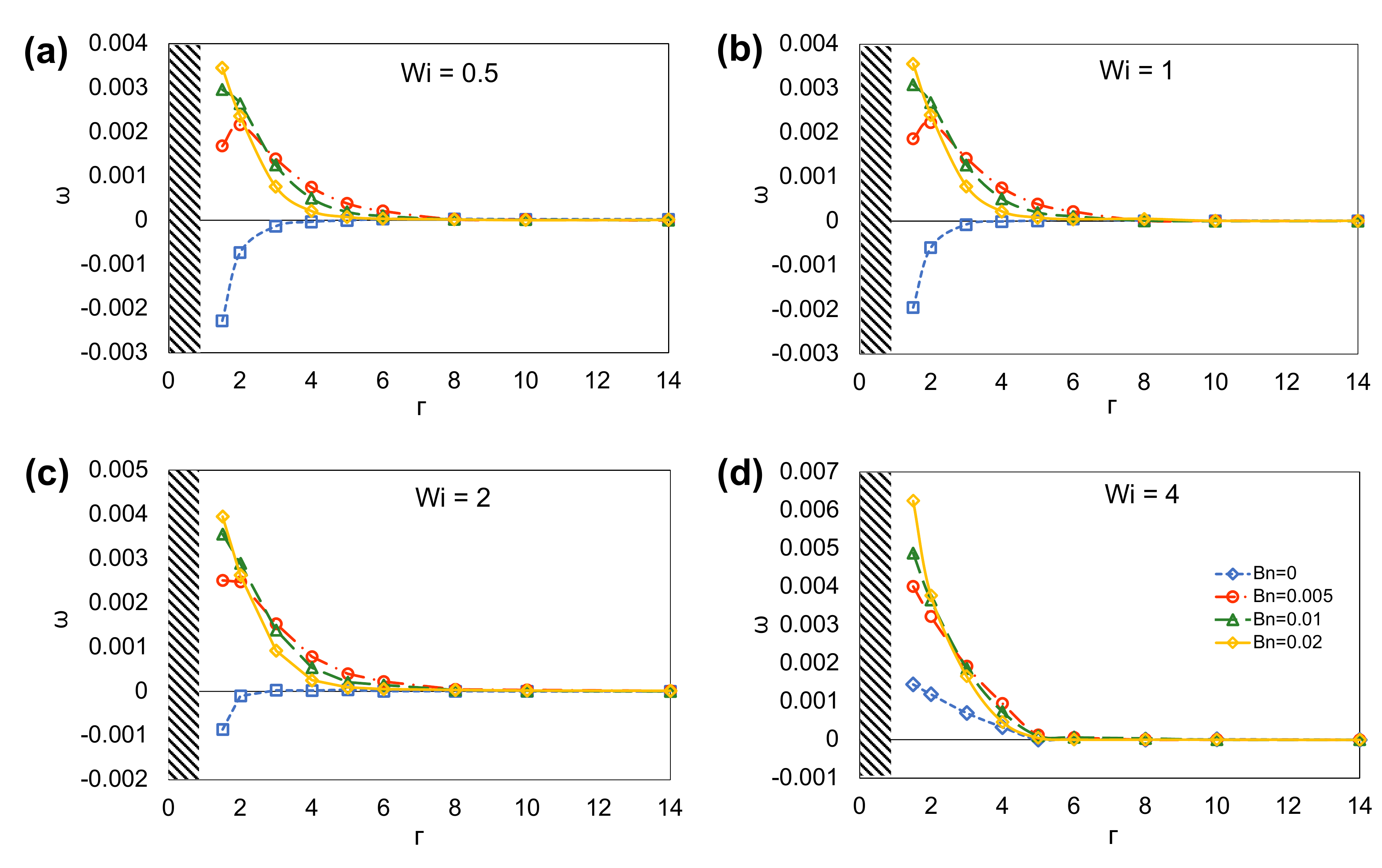}
\caption{Angular velocity of the particle as a function of the particle-wall distance for different values of the Weissenberg and Bingham number. The shaded area is the region inaccessible to the particle due to its finite size.} 
\label{fig:omega}
\end{figure}

Finally, we report in Fig.~\ref{fig:transient}c and \ref{fig:omega} the transient and mastercurves of the particle angular velocity. For a purely viscoelastic fluid at low $Wi-$values, the angular velocity is negative at any distance from the wall, denoting a clockwise rotation with respect to the $z-$direction, in agreement with the Newtonian case \cite{Happel1983low}. As the Weissenberg number increases, the angular velocity decreases and changes sign at $Wi=4$, as discussed in previous works \cite{Singh2000sedimentation, Liu1993anomalous}. For an elastoviscoplastic suspending fluid, the rotation is always counterclockwise. Notice also that the angular velocity becomes nearly zero at a particle-wall distance of about 5-6 radii, that is shorter as compared to the distance such that the migration velocity becomes nil or the settling velocity constant.

\begin{figure}[t!]
\centering
\includegraphics[width=1\textwidth]{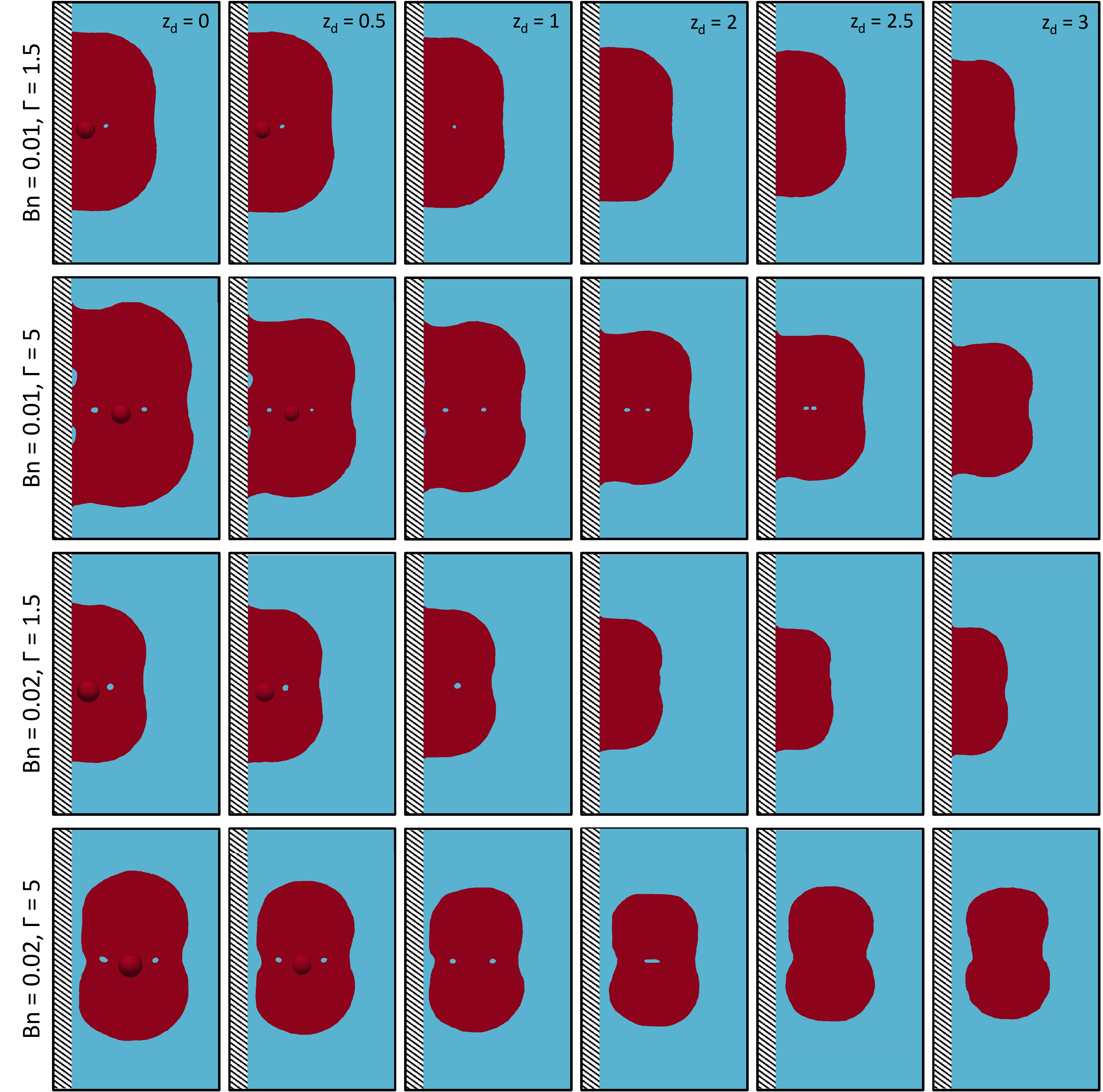}
\caption{Yielded (red) and unyielded (cyan) regions around the particle on slices parallel to the $xy-$plane at different dimensionless depth $z_\text{d}$. The slice at $z_\text{d}=0$ corresponds to the symmetry plane. Two values of the Bingham number and two particle-wall distances are considered. The Weissenberg number is $Wi=2$. The shaded area is the region inaccessible to the particle due to its finite size. The force is applied from top to bottom.}
\label{fig:yielded}
\end{figure}

An interesting feature of suspensions of particles in EVP fluids is the shape and extension of the yielded/unyielded zones around the particle, i.e., the regions in the fluid domain where $|\boldsymbol{\tau}_\text{d}|$ is higher (fluid behavior) or lower (solid behavior) than the yield stress. Figure~\ref{fig:yielded} displays the yielded (red) and unyielded (cyan) regions around the particle on some slices parallel to the symmetry $xy$-plane at different depths $z_\text{d}$ from such a plane. The Weissenberg number is fixed to $Wi=2$. The first and second rows refer to a particle-wall distance of $\Gamma=1.5$ and $\Gamma=5$, respectively, at a Bingham number of $Bn=0.01$. The same distances are set in the third and fourth rows, with the Bingham number increased to $Bn=0.02$.

A yielded zone surrounds the particle because of the high stress region due to the particle motion. As displayed in the leftmost panels (corresponding to the symmetry plane), at least two unyielded regions are visible in the examined cases. The first one is the external region surrounding the fluid zone. The second region is the small circle appearing at one or both sides of the particle. The existence of these isolated unyielded regions has been reported in Sabarian et al.~\cite{Sarabian2022interface} for the case of a sphere settling in an EVP material at the centerplane between two parallel walls and at different confinement ratios. However, as visible in Fig.~\ref{fig:yielded}, our results indicate that the small unyielded region in the gap disappears for a sufficiently low particle-wall distance. An inspection of the unyielded region shape inside the domain reveals that, at $\Gamma=5$, the two circles get closer and form a connected domain at $z_\text{d}\approx 2.5$ for $Bn=0.01$ and at $z_\text{d}\approx 2$ for $Bn=0.02$ (see the sequence of slices at increasing depth from the symmetry plane). Hence, the internal unyielded region is a ring surrounding the particle, as also found in Sabarian et al.~\cite{Sarabian2022interface}. On the other hand, at $\Gamma=1.5$, the unyielded region is an island with angular extension of about $30^\circ$ from the symmetry plane situated at the unconfined side of the particle. Indeed, by looking at the slices inside the domain, we observe that the circle disappears from $z_\text{d}=1$ to $z_\text{d}=2$ for both Bingham numbers. Sabarian et al.~\cite{Sarabian2022interface} found the ring  for all the considered confinement ratios. It should be mentioned that the lowest confinement ratio studied in their work corresponds to a particle-wall distance of $\Gamma=2$ and the two circles are visibly smaller than the cases of less confined systems. It is likely to assume that, by further reducing the confinement ratio, the ring disappears (probably leaving two isolated islands on the forth and back of the sphere). Two more unyielded regions are visible in the second row of Fig.~\ref{fig:yielded} close to the wall. These two regions are present only in a specific range of distances, provided that the particle is not too close to the wall and the boundary of the yielded region touches the wall. These regions are islands embedded in the yielded region, as confirmed by the sequence of slices for $Bn=0.01$ and $\Gamma=5$ showing that these two regions progressively reduce and disappear at about $z_\text{d}=1$. 

\begin{figure}[t!]
\centering
\includegraphics[width=0.5\textwidth]{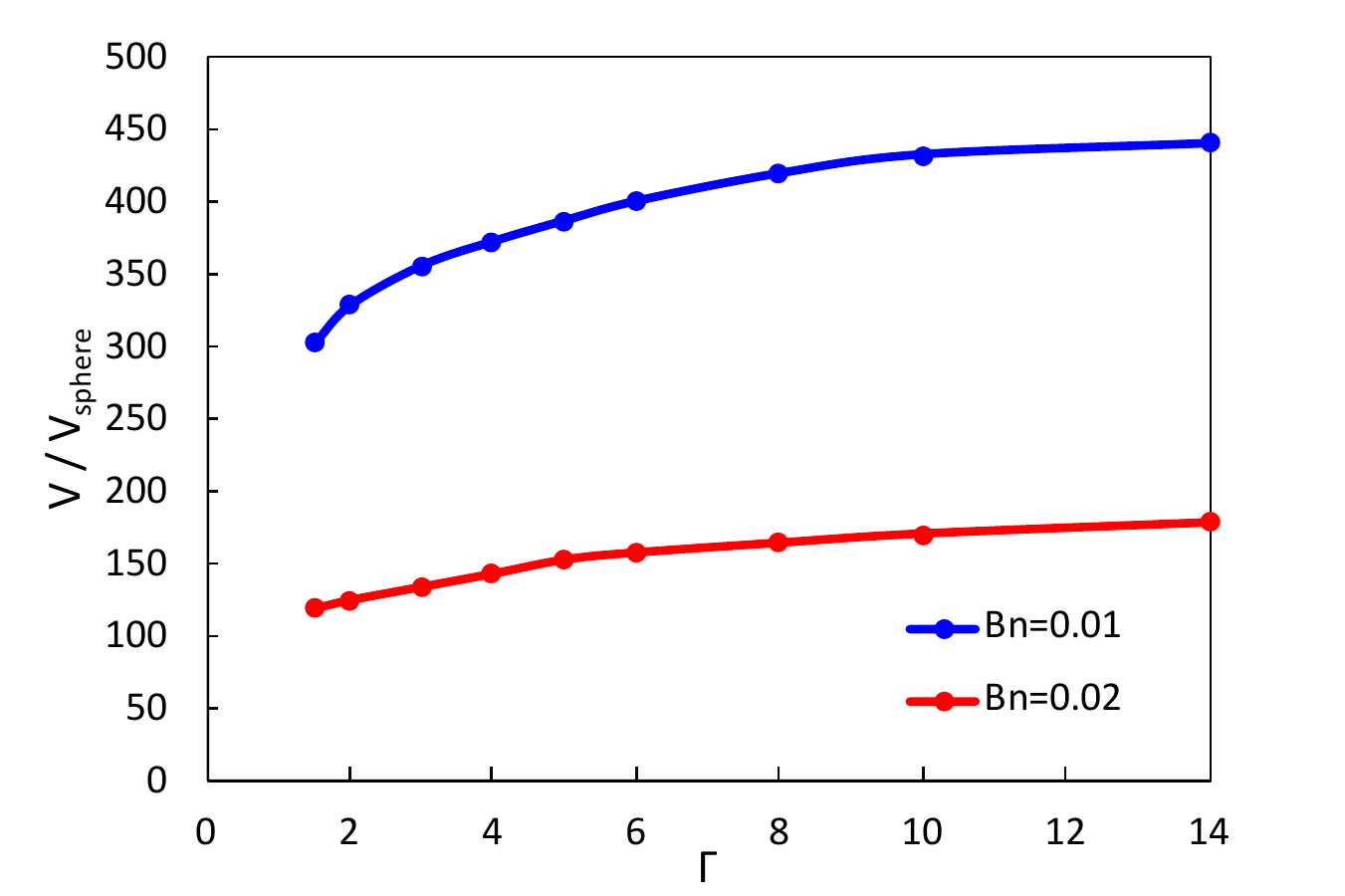}
\caption{Volume of the yielded region as a function of the particle-wall distance for $Bn=0.01$ and $Bn=0.02$. The volume is normalized by the particle volume.}
\label{fig:volume}
\end{figure}

As reported in the literature~\cite{Fraggedakis2016yielding,Sarabian2022interface}, the extension of the yielded region reduces by increasing the Bingham number. The presence of the wall does not significantly affect the extension of the unconfined side of the yielded region, which is also similar to the extension in the neutral $z-$direction. Finally, we show in Fig.~\ref{fig:volume} the volume of the yielded region, normalized by the particle volume, by varying the distance of the particle from the wall. As expected, the volume increases for increasing particle-wall distances up to reaching a constant value corresponding to the unconfined case, and a lower volume is found for the highest $Bn$-value.

Regarding the shape of the yielded region, some similarities can be noticed with the shapes reported in Sabarian et al.~\cite{Sarabian2022interface} for the confined cases. However, our results indicate a more symmetric shape in the fluid region above and below the sphere, whereas the top boundary of the yielded region reported in Sabarian et al.~\cite{Sarabian2022interface} and Fraggedakis et al.~\cite{Fraggedakis2016yielding} is closer to the particle.

\begin{figure}[t!]
\centering
\includegraphics[width=1\textwidth]{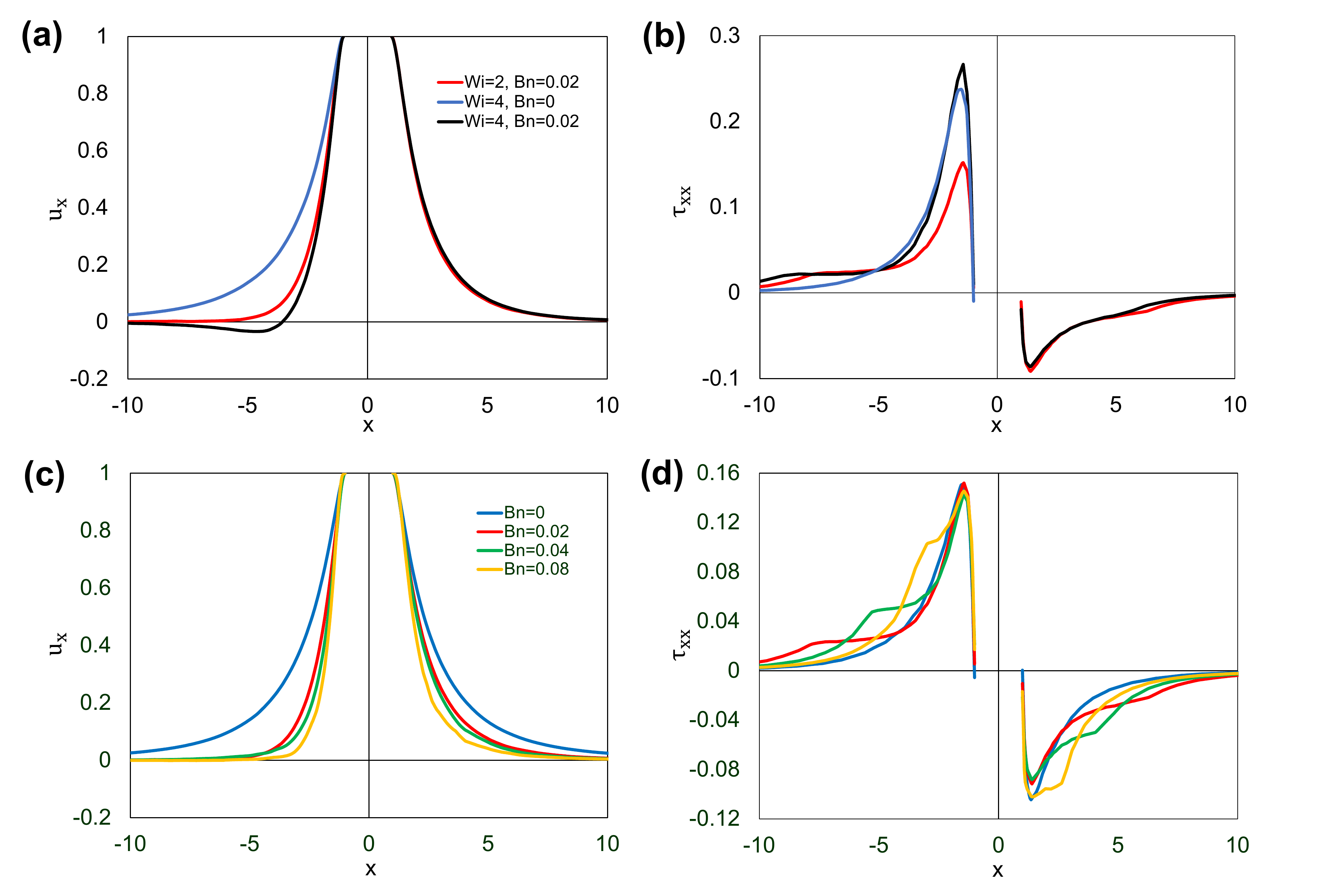}
\caption{Fluid velocity (left) and axial stress component (right) along the line passing through the particle center and parallel to the sedimentation direction for different values of the Weissenberg and Bingham numbers. The initial distance is $\Gamma=1.5$ and the profiles are taken after the initial transient due to the stress development. The fluid velocity is normalized by the sedimentation velocity of the particle taken at the same instant.} 
\label{fig:negativewake}
\end{figure}

Another interesting feature is the negative wake phenomenon behind the sphere. The blue curve in Fig.~\ref{fig:negativewake}a shows the component of the fluid velocity along the sedimentation direction, $u_\text{x}$, on a line passing through the center of the sphere and parallel to the $x-$direction for the purely viscoelastic case ($Bn=0$) at $\Gamma=1.5$ and $Wi=4$. In Fig.~\ref{fig:negativewake}b, the $xx-$component of the viscoelastic stress, $\tau_\text{xx}$, is reported. Both quantities are taken just after the initial transient due to the stress development. The fluid velocity is normalized by the sedimentation velocity of the particle taken at the same instant. The fluid velocity is always positive, denoting no negative wake behind the sphere. The red curves refer to the EVP fluid for $\Gamma=1.5$, $Wi=2$, and $Bn=0.02$. Despite the lower value of the Weissenberg number, the fluid velocity decays faster due to the plasticity effect. Indeed, the distribution of $\tau_\text{xx}$, qualitatively similar to that reported in Fraggedakis et al.~\cite{Fraggedakis2016yielding}, shows an abrupt change of the trends (at around $x=-8$) due to the transition from the yielded to the unyielded region. As discussed in Fraggedakis et al.~\cite{Fraggedakis2016yielding}, the residual normal stresses in the solid-like region pull the solid away from the sphere, possibly leading to the backflow motion, promoting the negative wake. Finally, by increasing the Weissenberg number to $Wi=4$ (black curves), the axial stress behind the sphere is both higher in magnitude and has a steeper slope as compared to the red curve that, together with the pulling effect of the solid-like region, leads to a even faster decay of the fluid velocity downstream with the appearance of the negative wake.

The effect of the Bingham number on the flow field behind the sphere is displayed in the bottom panels of Fig.~\ref{fig:negativewake}. The Weissenberg number fixed to $Wi=2$. The blue curve corresponds to the purely viscoelastic case, $Bn=0$, and the red one is the same of the top panels, i.e., for $Bn=0.02$. As the Bingham number increases, the abrupt change in $\tau_\text{xx}$ approaches the sphere as the yielded region surrounding the particle becomes smaller. The trends of the axial stress in the fluid layer just behind and in front of the sphere (i.e., from the maximum/minimum of $\tau_\text{xx}$ up to the particle surface) are, however, unaffected. Hence, both the magnitude and the gradient of the axial stress do not change as $Bn$ increases. Nevertheless, the fluid velocity decays faster at higher Bingham numbers due to the aforementioned `pulling effect' of the solid region. Although not clearly visible, the yellow curve Fig.~\ref{fig:negativewake}c, corresponding to $Bn=0.08$, is below the $x-$axis in a small region, denoting the onset of the negative wake phenomenon. Notice that the velocity decay is observed upstream the sphere too where a similar abrupt change of the axial stress occurs due to the transition from yielded to unyielded region. This is at variance with the purely viscoelastic case where a much weaker variation of the velocity field upstream the sphere is observed as the Weissenberg number is increased (see, e.g., Su et al.~\cite{Su2022data}).

\begin{figure}[t!]
\centering
\includegraphics[width=0.5\textwidth]{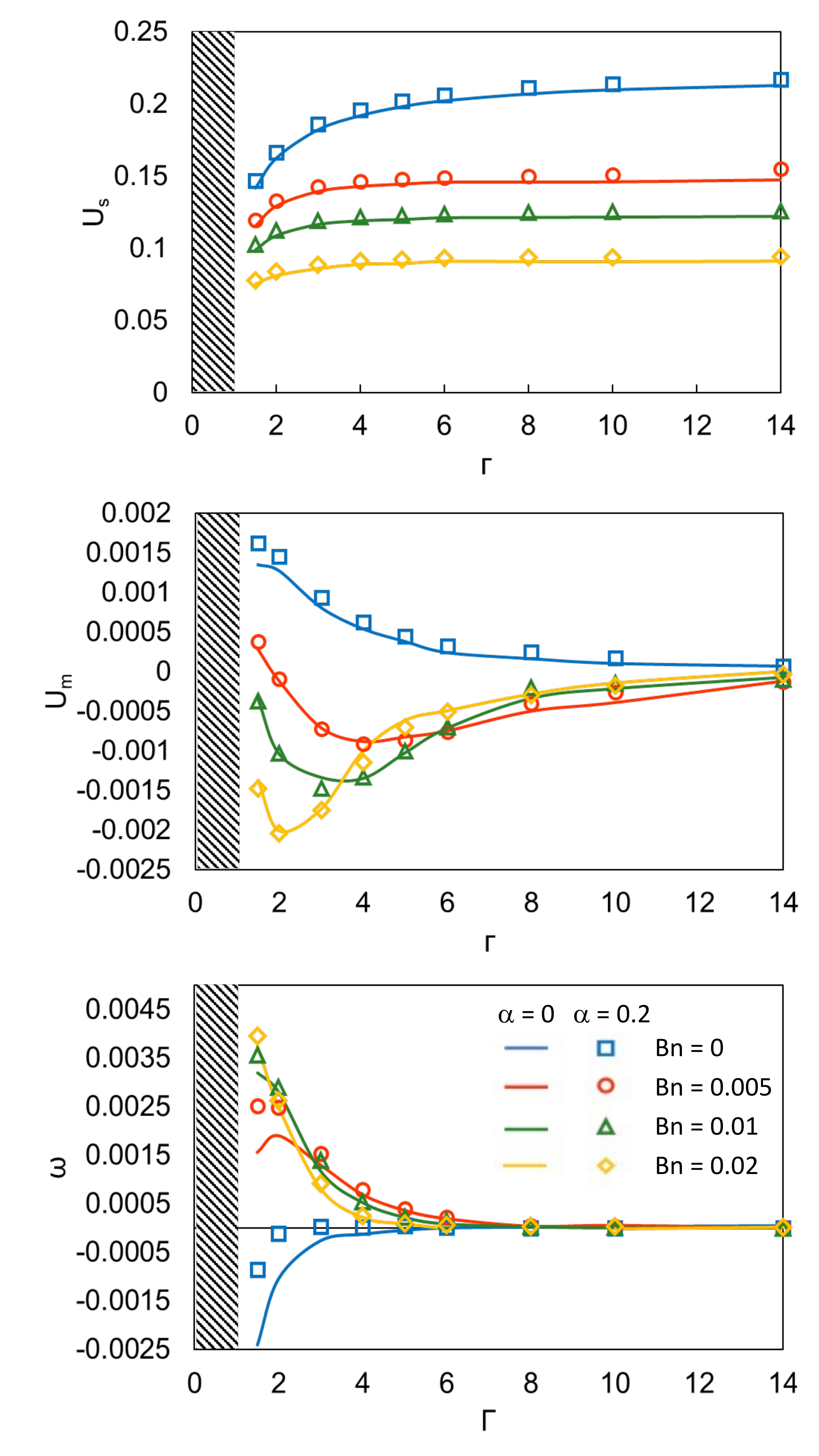}
\caption{Sedimentation, migration, and angular velocity of the particle as a function of the particle-wall distance for the different values of the Weissenberg and Bingham number. The solid lines refer to the Saramito model ($\alpha=0$) and the symbols to the Saramito-Giesekus model with $\alpha=0.2$. The shaded area is the region inaccessible to the particle due to its finite size.}
\label{fig:alpha0}
\end{figure}

We conclude this section by showing some simulation results at $\alpha=0$, i.e., for the original Saramito model. The sedimentation, migration, and angular particle velocities are reported in Fig.~\ref{fig:alpha0} for several particle-wall distances and $Wi=2$ as solid lines. For comparison, the same quantities at $\alpha=0.2$ are shown as symbols (the data are those already presented in the previous figures). It is readily observed that the sedimentation and migration velocities are very weakly affected by the mobility parameter. Specifically, only a slight decrease of these two quantities is found, in agreement with the purely viscoelastic case \cite{Villone2011numerical,Su2022data}. A more relevant quantitative deviation at $\alpha=0$ is observed for the angular velocity at small particle-wall distances and for low Bingham numbers. However, the trends and the sign of the angular velocity are the same as the case at $\alpha=0.2$. To justify the quantitative similar behavior of the particle dynamics between the Saramito and Saramito-Giesekus model, we evaluated the effective Weissenberg number, $Wi_\text{eff}=\lambda\dot{\gamma}$, in the fluid domain around the particle. It turns out that, even for the highest $Wi$-value investigated in this work, the largest effective Weissenberg number is lower than 1. Hence, as visible in the plots reported in the Appendix, a variation of the mobility parameter $\alpha$ does not produce any significant change in the rheological functions. The only difference between the two models is that the Giesekus variant predicts a non-zero second normal stress difference in shear flow. This, however, does not seem to have any relevant effect on the particle dynamics. As a final comment, we mention that, despite no relevant change is observed between $\alpha=0$ and $\alpha=0.2$, we experienced some numerical issues in the former case at $Bn=0$ and $Bn=0.005$, and for low particle-wall distances. Specifically, both the mesh and the time step needed to be refined to get convergent solutions. Recalling that the Saramito model predicts a divergent extensional viscosity (see Fig.~\ref{fig:app_elong}), numerical instabilities can be likely due to high elongational stress regions around the particle.

\section{Conclusions}

The dynamics of a rigid spherical particle in an elastoviscoplastic fluid in proximity of a flat wall subjected to a force parallel to the wall and under inertialess conditions is studied by direct finite element numerical simulations. The Giesekus-Saramito constitutive equation is employed to model the fluid. The Arbitrary Lagrangian-Eulerian formulation is used to handle the particle motion. The code is validated by simulating the settling dynamics of a sphere in an EVP fluid using the Saramito constitutive equation in an infinite medium. Mesh and time convergence are verified by comparing the time evolution of the particle kinematic quantities for different mesh resolutions and time step sizes.

The particle dynamics in terms of the translational and rotational velocities is studied for various particle-wall distances, Weissenberg and Bingham numbers. The purely viscoelastic case is also simulated to understand the effect of plasticity on the particle motion. The sedimentation slows down as the Weissenberg number decreases and the Bingham number increases. The mastercurves of the migration velocity indicate a lateral motion of the particle away from the wall in the purely viscoelastic case. However, as the Bingham number increases, the migration velocity progressively decreases and the particle inverts its migration direction and approaches to the wall. The inversion of the migration direction from the purely viscoelastic to the EVP case is due to a different shear rate distribution around the particle induced by the yield stress. In the EVP material, indeed, the shear rate gradient in the particle-wall gap reduces to the lower sedimentation speed and, at the same time, the presence of the unyielded region leads to a larger shear rate gradient at the particle side far from the wall, generating, in fact, a net force towards the wall.

Regarding the flow field, two unyielded regions are found: an external region surrounding the fluid zone around the particle, and isolated small regions at one or both sides of the particle. No negative wake is observed up to $Wi=2$ and $Bn=0.04$. However, as $Bn$ increases, the yielded region becomes smaller and the solid region approaches the particle. The residual normal stresses in this region pull the solid away from the particle, leading to a faster decay of the fluid velocity behind the sphere. Indeed, at $Bn=0.08$, a zone in the wake with inverted velocity is observed. Such a phenomenon is more pronounced at $Wi=4$. 

The results of this study show that the interplay of confinement, elasticity and plasticity significantly affects the settling dynamics of a spherical particle as well as the shape and extension of the yielded/unyielded regions. This suggests that the critical value of the Bingham number leading to the blockage of the particle depends on the confinement. Future work will address the role of the confining wall on the stopping criteria. Finally, the present work as well as those available in the literature are limited to spheres. The particle motion and flow dynamics become more complicated in the case of non-spherical particles as the orientation plays a relevant role. The next step will be the extension of the present study to anisotropic particle shapes suspended in EVP fluids.

\section*{Acknowledgements}

This work is carried out in the context of the project YIELDGAP (https://yieldgap-itn.eu) that has received funding from the European Union’s Horizon 2020 research and innovation programme under the Marie Skłodowska-Curie grant agreement No 955605.

\section*{Appendix}

\begin{figure}[t!]
\centering
\includegraphics[width=0.5\textwidth]{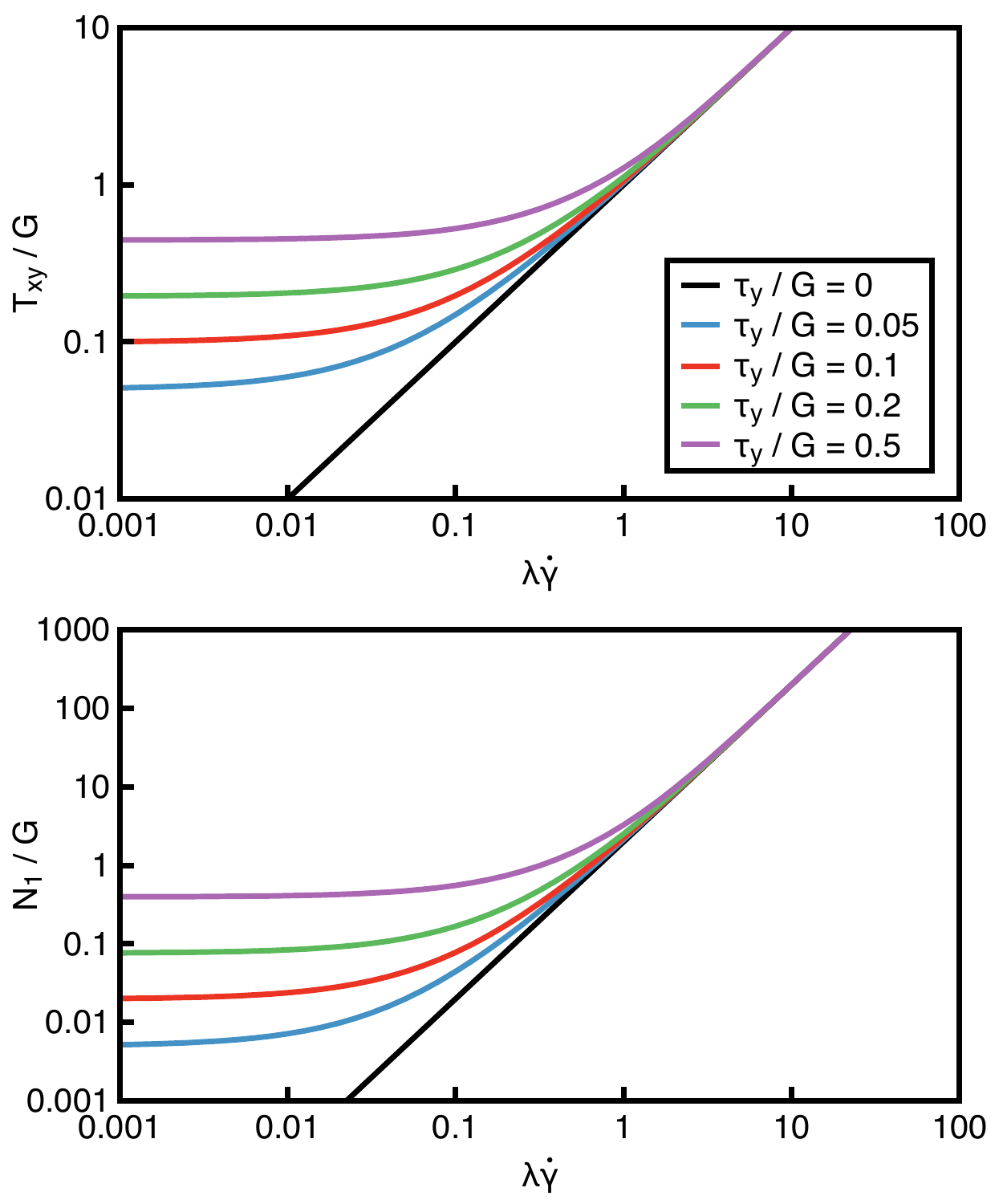}
\caption{Shear stress (top) and first normal stress difference (bottom) as a function of the Weissenberg number for the Saramito model in simple shear flow at different values of the yield stress. The stresses are made dimensionless with the elastic modulus $G=\eta_\text{p}/\lambda$.}
\label{fig:app_shear_alpha0}
\end{figure}

The rheological properties predicted by the Saramito-Giesekus model, Eq.~\eqref{eqn:Saramito_GSK}, under shear and uniaxial elongational flows are shown in this Appendix. For comparison, the same properties for the original Saramito model are reported as well. In all the cases, we do not consider the viscosity solvent contribution, i.e., $\eta_\text{s}=0$ in Eq.~\eqref{eqn:viscoelastic constitutive}. Figure~\ref{fig:app_shear_alpha0} displays the shear and normal stresses for the Saramito model at various values of the yield stress. Specifically, the black curve is for a zero yield stress, recovering the Upper-Convected Maxwell (UCM) equation. Notice that the stresses (including the yield stress) are made dimensionless by the elastic modulus $G=\eta_\text{p}/\lambda$ and the abscissa is the Weissenberg number defined as $\lambda \dot{\gamma}$.

\begin{figure}[t!]
\centering
\includegraphics[width=0.5\textwidth]{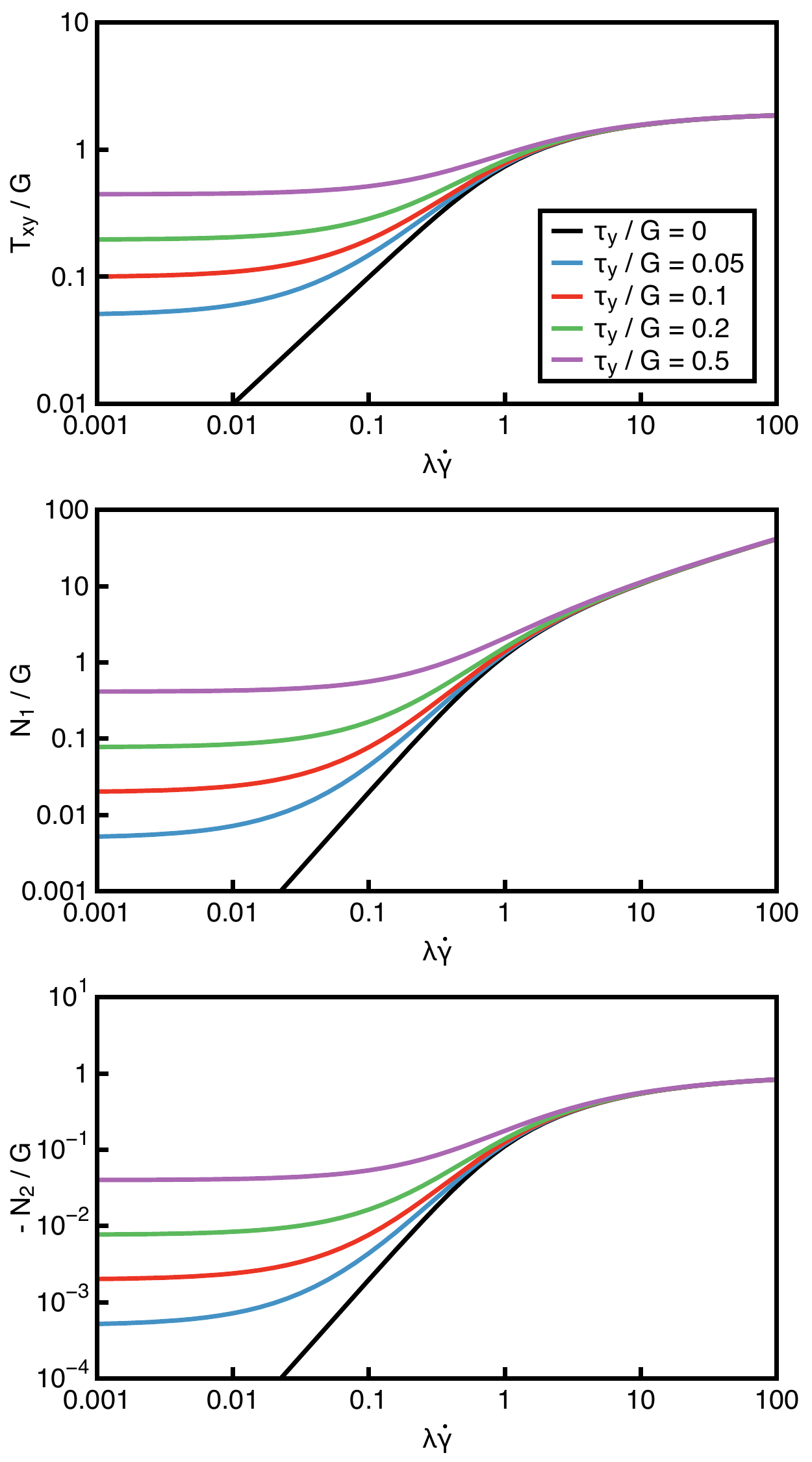}
\caption{Shear stress (top), first normal stress difference (middle), and second normal stress difference as a function of the Weissenberg number for the Saramito-Giesekus model in simple shear flow at different values of the yield stress. The mobility parameter is set to $\alpha=0.2$. The stresses are made dimensionless with the elastic modulus $G=\eta_\text{p}/\lambda$.} 
\label{fig:app_shear_alpha02}
\end{figure}

The same quantities for the Saramito-Giesekus model with $\alpha=0.2$ are shown in Fig.~\ref{fig:app_shear_alpha02}. The third panel reports the (negative values of the) second normal stress difference which is not zero in this constitutive equation (as for the original Giesekus model). Apart from the existence of a second normal stress difference, other differences from the Saramito model are the shear-thinning behavior of the viscosity and of the first and second normal stress difference coefficients (beyond a typical shear rate value, the shear stresses increase less than linear and the normal stresses less than quadratic), similarly to what observed by comparing the Giesekus and UCM models.

\begin{figure}[t!]
\centering
\includegraphics[width=1\textwidth]{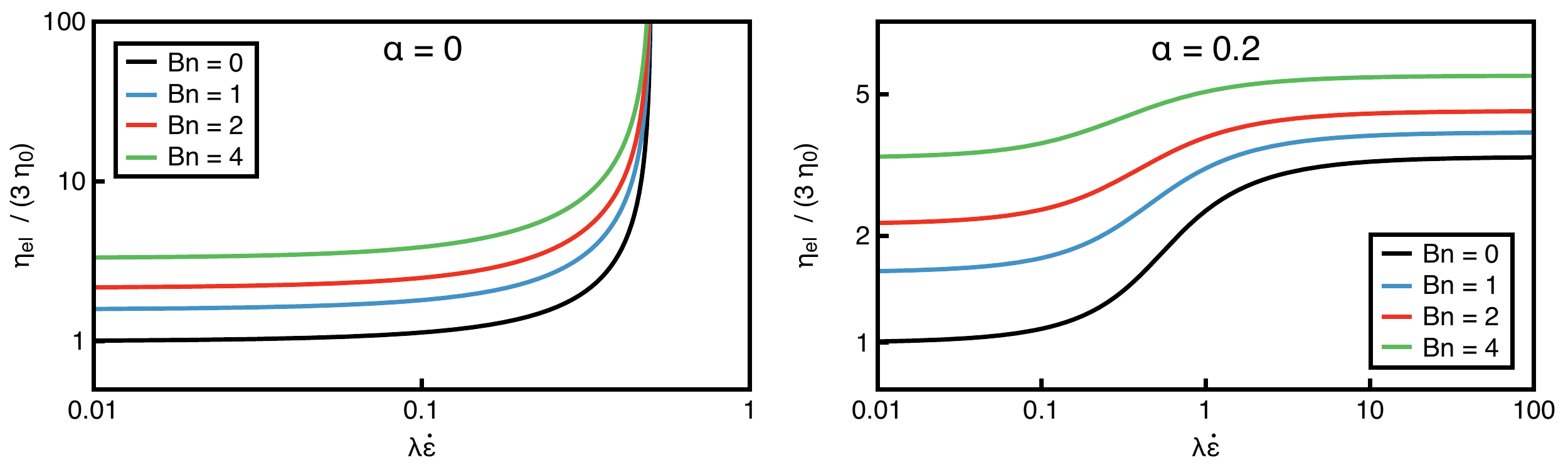}
\caption{Extensional viscosity for the Saramito (left) and Saramito-Giesekus (right) models in uniaxial extensional flow as a function of the Weissenberg number for different values of the Bingham number. The mobility parameter for the Saramito-Giesekus model is set to $\alpha=0.2$. The extensional viscosity is normalized by $3\eta_0.$} 
\label{fig:app_elong}
\end{figure}

Finally, in Fig.~\ref{fig:app_elong}, the uniaxial extensional viscosity normalized by $3\eta_0$ is shown as a function of the Weissenberg number defined as $\lambda \dot{\epsilon}$ with $\dot{\epsilon}$ the elongational rate. Curves at different Bingham numbers, defined as $Bn=\tau_\text{y}/(\eta_0 \dot{\epsilon})$, are reported. The left and right panels refer to the Saramito and Saramito-Giesekus model with $\alpha=0.2$. As it is well-known \cite{Saramito2007new}, the Saramito model predicts a divergent extensional viscosity at $Wi=0.5$ (as for the UCM model), regardless of the value of the Bingham number. The Giesekus variant has a bounded extensional viscosity (as for the Giesekus model) at high Weissenberg numbers with an horizontal asymptote depending on the Bingham number. As a final remark, we point out that the rheological predictions for the Saramito-Giesekus model are qualitatively similar to those of the Phan-Thien-Tanner (PTT) variant of the Saramito model, proposed in the original paper \cite{Saramito2007new}, with the only difference being $N_2=0$ in the Saramito-PTT (just like the original PTT model).

\bibliography{bibliography.bib}

\end{document}